\documentclass{osa-article}
\journal{oe}

\articletype{Research Article}

\usepackage{lineno}
\usepackage{subcaption} 
\usepackage{adjustbox}  
\usepackage{amsmath}
\usepackage{xcolor}
\usepackage{bm}
\usepackage{boldline}
\usepackage{soul}
\usepackage[title]{appendix}

\DeclareMathOperator*{\argmin}{arg\,min}

\newcommand{\genparams}{{\theta_g}}
\newcommand{\discparams}{{\theta_d}}
\newcommand{\disc}{D_{\discparams}} 
\newcommand{\gen}{G_{\genparams}} 

\graphicspath{{plots/},{plots/mh_figures/}}
\newcommand{\um}{{\rm \mu m}}
\newcommand{\cm}{{\rm cm}}
\newcommand{\gr}{{\rm g}}

\newcommand{\mps}{{\rm m}\;{\rm s}^{-1}}
\newcommand{\gcc}{{\rm cm}\;{\rm g}^{-3}}

\begin{document}

\title{High-Precision Inversion of Dynamic Radiography Using Hydrodynamic Features}
\author{Maliha Hossain,\authormark{1,*} 
  Balasubramanya~T.~Nadiga,\authormark{2}
  Oleg~Korobkin,\authormark{2}
  Marc~L.~Klasky,\authormark{2}
  Jennifer~L.~Schei,\authormark{2}
  Joshua~W.~Burby,\authormark{2}
  Michael~T.~McCann,\authormark{2}
  Trevor Wilcox,\authormark{2}
  Soumi~De,\authormark{2} and Charles~A.~Bouman\authormark{1}}

\address{\authormark{1}Purdue University, University, West Lafayette, IN 47907, USA\\
\authormark{2}Los Alamos National Laboratory, Los Alamos, NM 87545, USA}

\email{\authormark{*}mhossain@purdue.edu} 



\begin{abstract}
Radiography is often used to probe complex, evolving density fields in dynamic systems and in so doing gain insight into the underlying physics. It has a long history dating back to Radon. This technique has been used in a number of fields including materials science, shock physics, inertial confinement fusion, and other national  security applications.  In many of these applications, however, complications resulting from noise, scatter, complex beam dynamics, etc. prevent the reconstruction of density from being accurate enough to identify the underlying physics with sufficient confidence.  As such, density reconstruction from static/dynamic radiography has typically been limited to identifying discontinuous features such as cracks and voids in a number of these applications.

In this work, we propose a fundamentally new approach to reconstructing density from a temporal sequence of  radiographic images. Using only the robust features identifiable in radiographs, we combine them with the underlying  hydrodynamic equations of motion using a machine learning approach---conditional generative adversarial network (cGAN)---to determine the density fields from a dynamic sequence of radiographs.
Next, we seek to further enhance the hydrodynamic consistency of the ML-based density reconstruction through a process of parameter estimation and projection onto a hydrodynamic manifold.  In this context, we note that the distance from the hydrodynamic manifold given by the training data to the test data in the parameter space considered both serves as a diagnostic of the robustness of the predictions and serves to augment the training database, with the expectation that the latter will further reduce future density reconstruction errors.  Finally, we demonstrate the ability of this method to outperform a traditional radiographic reconstruction in capturing allowable hydrodynamic paths even when relatively small amounts of scatter is present. 
\end{abstract}

\section{Introduction}
\label{sec:intro}

The reconstruction of a density object from  line integrated radiographic projections has a long history dating back to Radon~\cite{radon1917}.  Accordingly, in many scientific applications arising in material science, shock physics, inertial confinement fusion, and in national security applications including stockpile stewardship, radiography plays a major role in probing complex, evolving density fields in dynamic systems as a means to understand the physics of the system.  However, density reconstructions performed  using experimental radiographic data of dynamic tests are complicated by the noisy and complex multi-scale and multi-physics environment and quantitative density evaluations  are a challenge.

Over the past several decades image reconstruction methods have evolved from simple analytical methods such filtered-back projection (FBP) methods, X-ray CT (e.g. Feldkamp-Davis-Kress or FDK methods), and the Inverse Abel Transform for axisymmetric systems \cite{cormack1963representation,feldkamp:84:pcb,bracewell1986fourier}.  These methods rely on relatively simple mathematical models of the imaging system. Modern approaches to solving the radiographic inversion problem for poly-energetic radiographic systems work with complex non-linear, non-convex forward models and employ iterative reconstruction techniques~\cite{elbakri_statistical_2002}.  Iterative reconstruction algorithms are based on more sophisticated models for the imaging system's physics and models for sensor and noise statistics.  These methods are often called model based image reconstruction (MBIR) methods or statistical image reconstruction (SIR) methods \cite{ravishankar2019image}. 

Mathematically, we may express the image reconstruction problem as:
\begin{equation}
\hat{x} = \argmin_x A(x) + \sum_i \alpha_i R_i(x)
\label{eq:opt_recon}
\end{equation}
where $\hat{x}$ denotes the reconstructed density profile, $A(x)$ is the data fidelity term capturing the forward model of the imaging process and the statistical models of measurements and noise, and the $R_i(x)$ are regularizer models for the desired image.  The scalar parameters $\alpha_i$ control the relative strength of the respective regularizers.  

\begin{figure}[!h]
  \centering
  \includegraphics[width=0.8\textwidth]{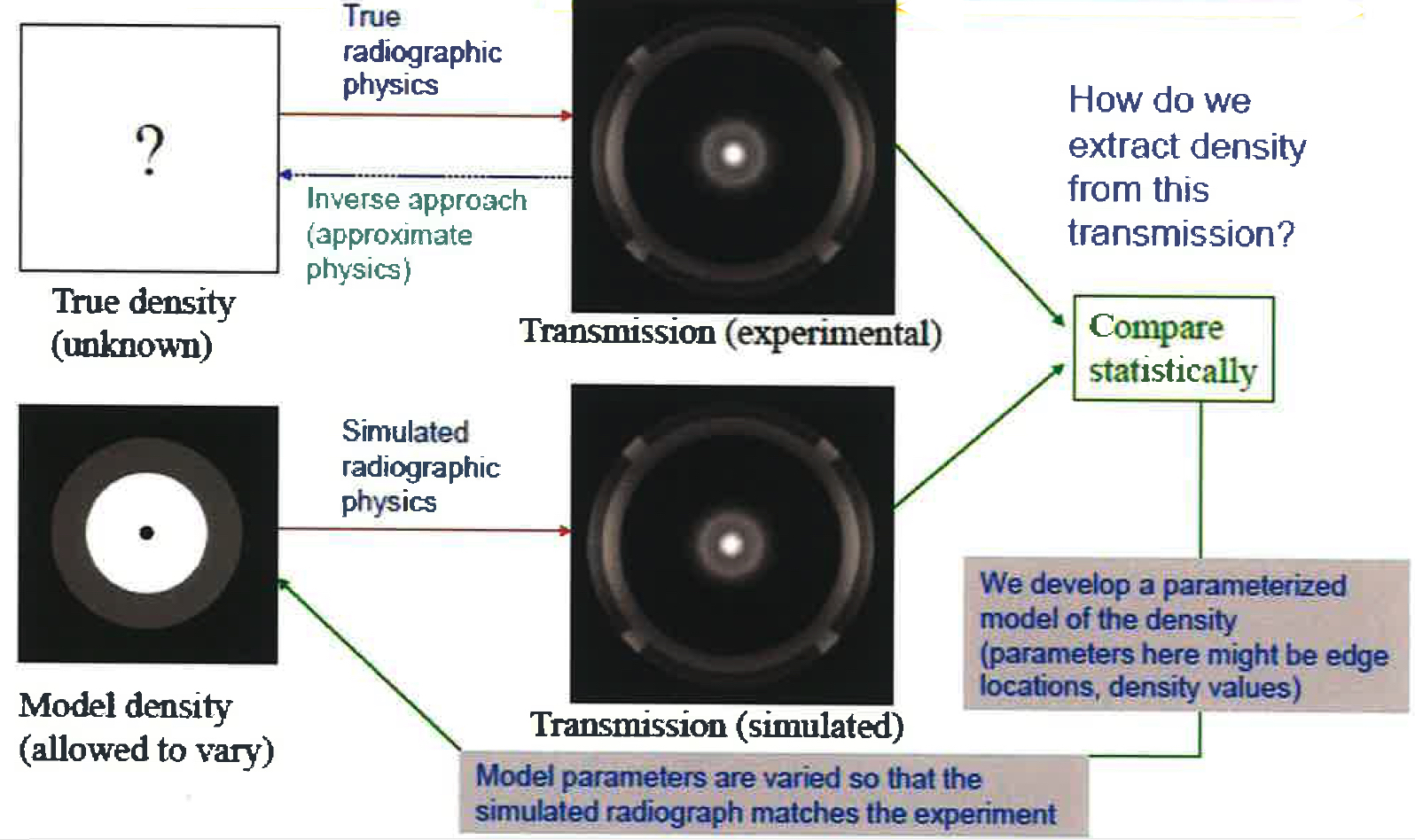}
  \caption{Forward modeling approach to radiographic reconstruction.}
  \label{fig:forward_model.jpg}
\end{figure}

Figure \ref{fig:forward_model.jpg}  depicts the objective of the forward modeling approach.   These  algorithms iteratively estimate the unknown image based on the system (physical or forward) model, measurement statistical model, and assumed prior information about the inspected object. For example,  minimizing  penalized  weighted-least  squares (PWLS) cost functions has been popular in  X-ray  CT,  and  these  costs  include  a  statistically weighted  quadratic  data-fidelity  term  (capturing  the  imaging forward model and noise variance) and a penalty term called a regularizer that models the prior information about the object.\cite{ravishankar2019image} 

The determination of the correct density distribution from \eqref{eq:opt_recon} is contingent not only upon formulating a proper physics-based forward model, but also on imposing appropriate regularization and effective   schemes to solve the optimization problem.

The uncertainties in the reconstruction process arise from our inability to exactly represent various aspects of the radiographic measurement system, such as scatter, beam spot movement, beam-target interactions, beam dynamics repeatability, and aspects of the image formation process.

Indeed, one of the major sources of error in the density reconstructions is attributed to scattered radiation.
Scatter typically creates a loss of contrast and leads to image artifacts such as cupping, shading, streaks, etc.
Scatter is caused by many types of photon-matter interactions~\cite{cohentannoudji_atom_1998} including Compton scatter, Rayleigh scatter, pair production, scatter involving the scene/background, etc.
Many techniques have been proposed for scatter correction~\cite{stonestrom_scatter_1976,sun_improved_2010,bhatia_convolution_2017,tisseur_evaluation_2018,maier_deep_2018,mccann_local_2021} (cf.~\cite{ruehrnschopf_general_2011,ruehrnschopf_general_2011a} for detailed review) in medical imaging, nondestructive testing, and other settings. 
The more recent techniques~\cite{maier_deep_2018,mccann_local_2021}
learn or fit scatter models based on training data including those generated from Monte Carlo N-particle transport code (MCNP) simulations~\cite{werner_mcnp6.2_2018}.

\begin{figure}[!h]
 \centering
 \includegraphics[width=0.95\textwidth]{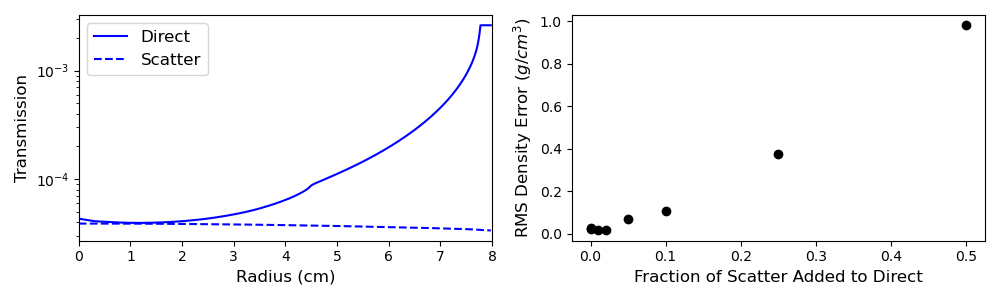}
 \begin{tabular}{cc}
  \includegraphics[width=0.48\textwidth]{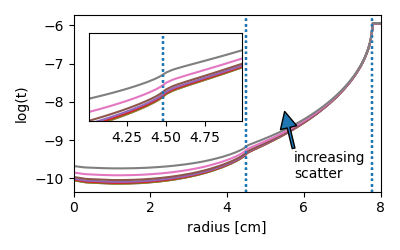}
  \includegraphics[width=0.48\textwidth]{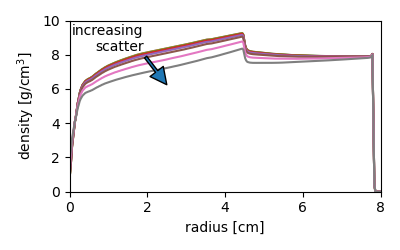}%
 \end{tabular}
 \caption{
    Top left panel: transmission of the direct (solid) and scattered (dashed 
    line) radiation through a sample object. 
    Top right panel: the RMS error $||\delta\rho||_2$ between the descattered 
    density and the ground truth, as a function of the fraction of scattered 
    radiation added to the direct transmission.
    Bottom left: transmission for different added levels of scatter. 
    Bottom right: descattered density profiles for the same levels of scatter.
    In transmission (bottom left), even in the presence of scatter, signatures
    of the hydrodynamic shock and edge (dotted lines) are clearly visible in 
    the radiograph and are not displaced by the scatter (see the inset).
    For the density profiles (right), on the other hand, significant distortions 
    are present after the density reconstruction, so that the RMS density error 
    (top right) is high.
 } 
  \label{fig:recon_and_edges_with_scatter}
\end{figure}

The complex nature of scatter makes it impossible to accurately capture even with sophisticated transport simulations, e.g.  MCNP simulations. The presence of anomalous scatter fields in experimental data, such as from scene scatter exacerbates the problem.
Such fields may not be accurately characterized and removed.   
Moreover, existing scatter (or other noise) correction techniques are not yet able to provide highly accurate quantitative density reconstructions~\cite{mccann_local_2021} and break down in the presence of anomalous fields including scatter attributed to surrounding structures. 
As an illustration we examined the propagation of a shock into a non-uniform density field created by the implosion of a steel shell.
A shock is a type of propagating disturbance that moves faster than the local speed of sound in the medium.  In our problem the shock is generated by the convergence of the steel shell at the axis.
Figure \ref{fig:recon_and_edges_with_scatter} in the upper left panel shows the relative levels of direct and scattered radiation, as calculated by MCNP \cite{werner_mcnp6.2_2018} for a density field generated from a hydrodynamic simulation using CTH \cite{hertel98b}.  

While the use of the traditional forward modeling approach to obtain quantitative density  fields has been demonstrated to be sensitive to the ability of the forward model to accurately capture scatter and other radiographic imaging effects, other aspects of the radiograph are relatively insensitive to these effects.
Figure ~\ref{fig:recon_and_edges_with_scatter}, bottom left panel, depicts the synthetic radiograph for a density field. 
Dotted lines indicate the hydrodynamic features, namely shock and edge, easily extracted using traditional edge finding techniques from synthetic radiographs, such as e.g. Canny Edge detection.
They are not significantly altered by the presence of the scattered field.
Figure~\ref{fig:recon_and_edges_with_scatter} also shows the effect of varying levels of scatter on density reconstruction. 
The apparent insensitivity of the hydrodynamic feature's localization to the scatter level indicates the robustness of these features.  
Uncertainties in edge location arise from experimental components such as beam motion, magnification error, detector resolution, and image noise; however, density reconstructions based upon feature location, including errors, may provide a higher degree of accuracy compared to density reconstructions based upon a forward model or inversion of a de-scattered radiographic image.

Figure~\ref{fig:recon_and_edges_with_scatter} supports an approach of focusing on the robust temporal sequence of shock and edge hydrodynamic features to perform density reconstructions. Accordingly, in this work, we explore the possibility that distinguished radiographic features, together with constraints imposed by dynamics, are sufficient by themselves to perform density reconstruction from radiographic sequences. We present a machine learning approach employing a conditional generative adversarial network (cGAN) to learn the mapping from radiographic features to the underlying density field. 

This approach seeks to obtain density profiles that lie on a hydrodynamic manifold rather than the traditional approach of using individual radiographic time slices to perform independent density reconstructions.  
Namely, when employing dynamic radiography, the density being imaged evolves from one snapshot to the next, leading to a sequence of correlated radiographs. These correlations may be explained using models from continuum mechanics such as hydrodynamics or elastodynamics. Owing to these correlations, more information may be gleaned from a radiographic sequence than from the individual images regarded as independent observations. 

The notion of a hydrodynamic manifold is a useful way to formalize the additional dynamical information present in a radiographic sequence of $N$ snapshots when compared with $N$ independent radiographs.
The ideal Euler equations comprise an example continuum model governing density evolution in many realistic scenarios where dynamic radiography might be applied. 

Assuming that appropriate boundary conditions are known and fixed, 
there is a well-defined mapping $\mathcal{H}:\sigma\mapsto z$, from tuples of the form 
\[
\sigma\equiv (P,\rho_0,\bm{u}_0,e_0),
\]
to solutions of the Euler equations, 
\[z\equiv (\rho,\bm{u},e),\]
where $\rho$ is the mass density, $\bm{u}$ is the fluid velocity, $e$ is the specific internal energy, and $p = P(\rho,e)$ is the pressure expressed as a point-wise function of $\rho$ and $e$. 
Here $P$ denotes an equation of state and $(\rho_0,\bm{u}_0,e_0)$ denotes an initial condition for the Euler equations. Note that $\rho_0,\bm{u}_0,$ and $e_0$ are functions of space alone, while $\rho,\bm{u},$ and $e$ are functions of both space and time. In full generality, the \textbf{hydrodynamic manifold} $\mathcal{M}$ is the image of 
hydrodynamic solution mapping $\mathcal{H}$. Note that $\mathcal{M}$ is a subset of a large function space containing arbitrary time-dependent density, velocity, and specific internal energy fields that satisfy the given boundary conditions. Among that most general class of fields, those contained in the hydrodynamic manifold are special because they solve the system of Euler equations, with \emph{some} equation of state, and \emph{some} initial condition.

The image of the space of physically admissible parameters $\Sigma$ under the hydrodynamic solution map $\mathcal{H}$ is the \textbf{restricted hydrodynamic manifold}, $\mathcal{M}_{\Sigma}\equiv \mathcal{H}(\Sigma)$. When the choice of $\Sigma$ is understood from context, we will abuse terminology and simply refer to $\mathcal{M}_{\Sigma}$ as ``the" hydrodynamic manifold. In fact, the restricted hydrodynamic manifold $\mathcal{M}_{\Sigma}$ is a smaller set than the hydrodynamic manifold $\mathcal{M}$, $\mathcal{M}_{\Sigma}\subset \mathcal{M}$. But $\mathcal{M}_{\Sigma}$ is physically more meaningful since it excludes solutions of the Euler equations (or whatever continuum model is relevant to a specific problem) that could never occur in an experiment anyway.

In this article, our primary interest in hydrodynamic manifolds lies in their ability to distinguish arbitrary sequences of putative density profiles from sequences that are consistent with dynamics. We will say that a sequence of candidate densities $\{\rho_i\}_{i=1,\dots,N}$, separated temporally by an increment $\Delta t$, \textbf{lies on the hydrodynamic manifold} if there is some $\sigma\in\Sigma$ and some $t_0\in\mathbb{R}$ such that $(\rho,\bm{u},e)=\mathcal{H}(\sigma)$ satisfies
\[
\rho(\bm{x},t_0+[i-1]\Delta t) = \rho_i(\bm{x}),\quad i=1,\dots,N.
\]
If a sequence does not lie on the hydrodynamic manifold, it is \textbf{dynamically-inconsistent}. Since the hydrodynamic manifold is a reflection of prior physics knowledge, reconstructions of dynamic radiographs that lie on a hydrodynamic manifold should be preferred over reconstructions that do not. 

This will enforce a physics-based constraint on the temporal evolution of the density field. To guarantee that the sequence of density fields obtained from the hydrodynamic features lie on a hydrodynamic manifold, we use  parameter estimation, if a parametric form is available or potentially learned, to project our solutions onto an available hydrodynamic manifold. 

In the remainder of this work we present details  of our new approach for performing density reconstructions using the robust features extracted from the dynamic radiographic images.
We introduce the dynamic radiography problem in Section \ref{sec:dynamic_radiography_backround}. 

Section \ref{sec:data-driven} presents two machine learning approaches using cGANs to allow for the determination of the density fields from a dynamic radiographic sequence from the robust hydrodynamic sequence. The results of the density reconstructions using the cGANs  are presented in \ref{sec:results} along with projections of the cGAN predictions to the hydrodynamic manifold.  Section \ref{sec:Discussion and Conclusion} presents a discussion of the observations and conclusions drawn from the numerical experiments.

\section{Dynamic radiography}
\label{sec:dynamic_radiography_backround}
\begin{figure}[!h]
    \centering
    \includegraphics[width=\textwidth]{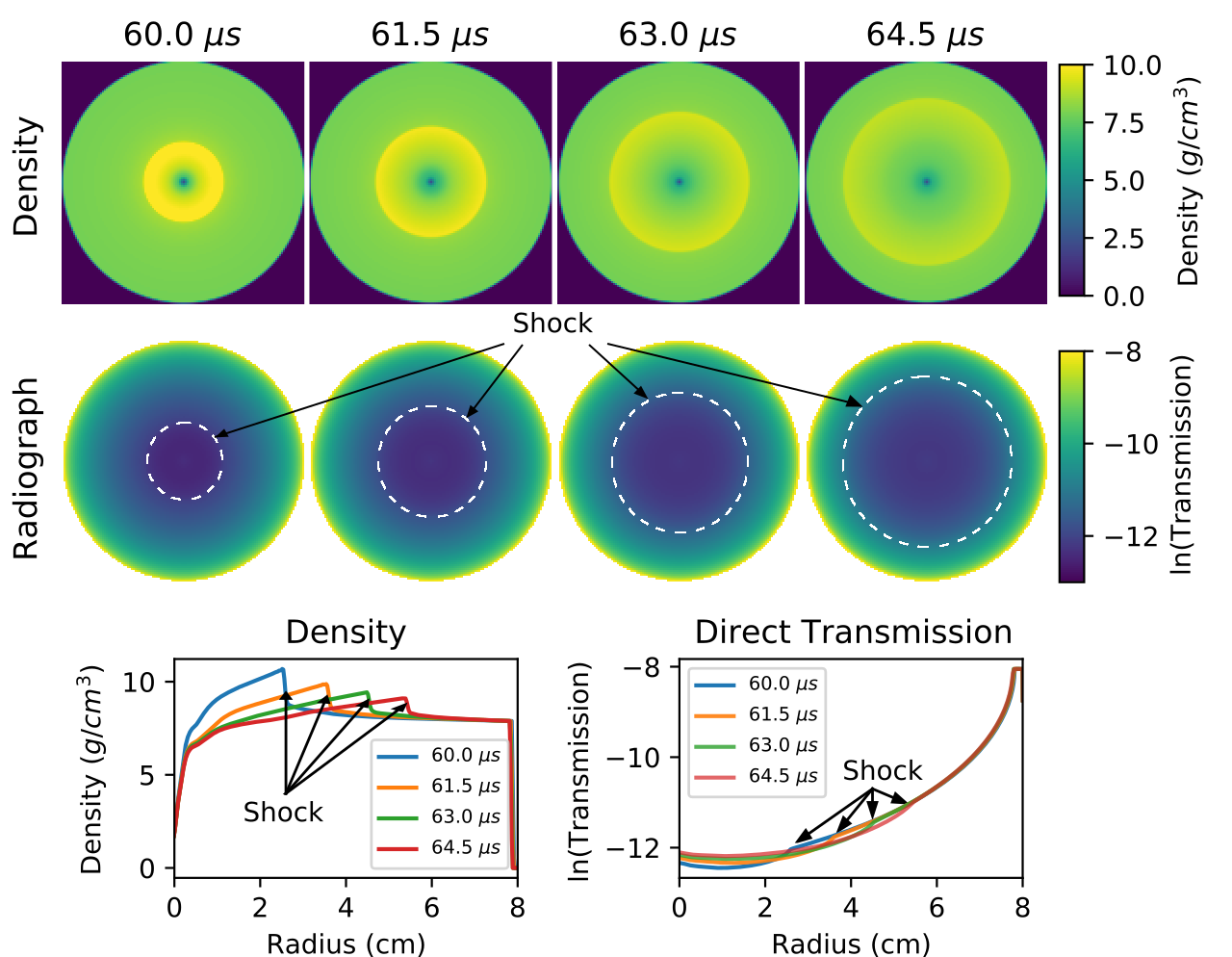}
    \caption{Density fields and total transmission radiographs from a series of four 1D hydrodynamic simulations.}
    \label{fig:density_direct_2D}
\end{figure}

A typical limited-view dynamic radiography problem may be described as follows. An experimenter conducts an experiment during which an object with density $\rho(\bm{x},t)$ evolves in time. Here $\bm{x}\in Q$ is a point in the experimental chamber $Q$ and $t\in [T_1,T_2]$ denotes time chosen from the interval $[T_1,T_2]$ over which the experiment takes place. At some instant $t_0\in[T_1,T_2]$, a sequence of $N$ X-ray pulses with temporal spacing $\Delta t$ begins to impinge on the object. Each pulse passes through the object, suffering attenuation determined by the optical depth of the target material. Upon emerging from the object, and potentially scattering off obstacles peripheral to the experiment, the attenuated pulses interact with the detector, producing one in a sequence of radiographic images. While the detector positions and orientations are known, small uncertainties in propagation characteristics of X-rays through the material (e.g. absorption cross sections) may produce significant uncertainties in modeled transmission amplitudes. Moreover, details of environmental scattered radiation are difficult or even impossible to describe from first principles. Additional radiographic issues that impede the ability to perform accurate density reconstructions include beam spot movement, beam energy spectrum fluctuations, beam-target interactions, and beam dynamics repeatability. The problem then is to determine the density field at the times $t_i = t_0 + [i-1]\Delta t$, $i=1,\dots,N$, given the radiographic sequences.  Figure \ref{fig:density_direct_2D} presents a sequence of density fields evolving in time along with the accompanying radiographic images.  The focus of our investigations will be on the imaging of axisymmetric objects, hence a single view of the object is sufficient to perform the density reconstruction if a perfect forward model were known.

While there have been some attempts to incorporate dynamic constraints in the radiographic community, ~\cite{robinson2003fast,cao2000time,ma11081395,sikdar2014dynamic,nemirovsky2011tomographic,krimerman2013reconstruction,gregson2012stochastic,ihrke2004image} most established techniques for dynamic radiographic X-Ray reconstruction  fail to exploit knowledge of laws underlying the time evolution of a density field to enhance accuracy. This dynamical information that relates different images in a radiographic sequence places constraints on candidate density fields that analysis of each radiograph in isolation cannot detect. It is therefore reasonable to suspect that new, more performant reconstruction techniques may be devised by incorporating these constraints.

\section{Test problem}
\label{sec:test_problem}

\begin{figure}[htbp]
  \centering
  \includegraphics[width=0.6\textwidth]{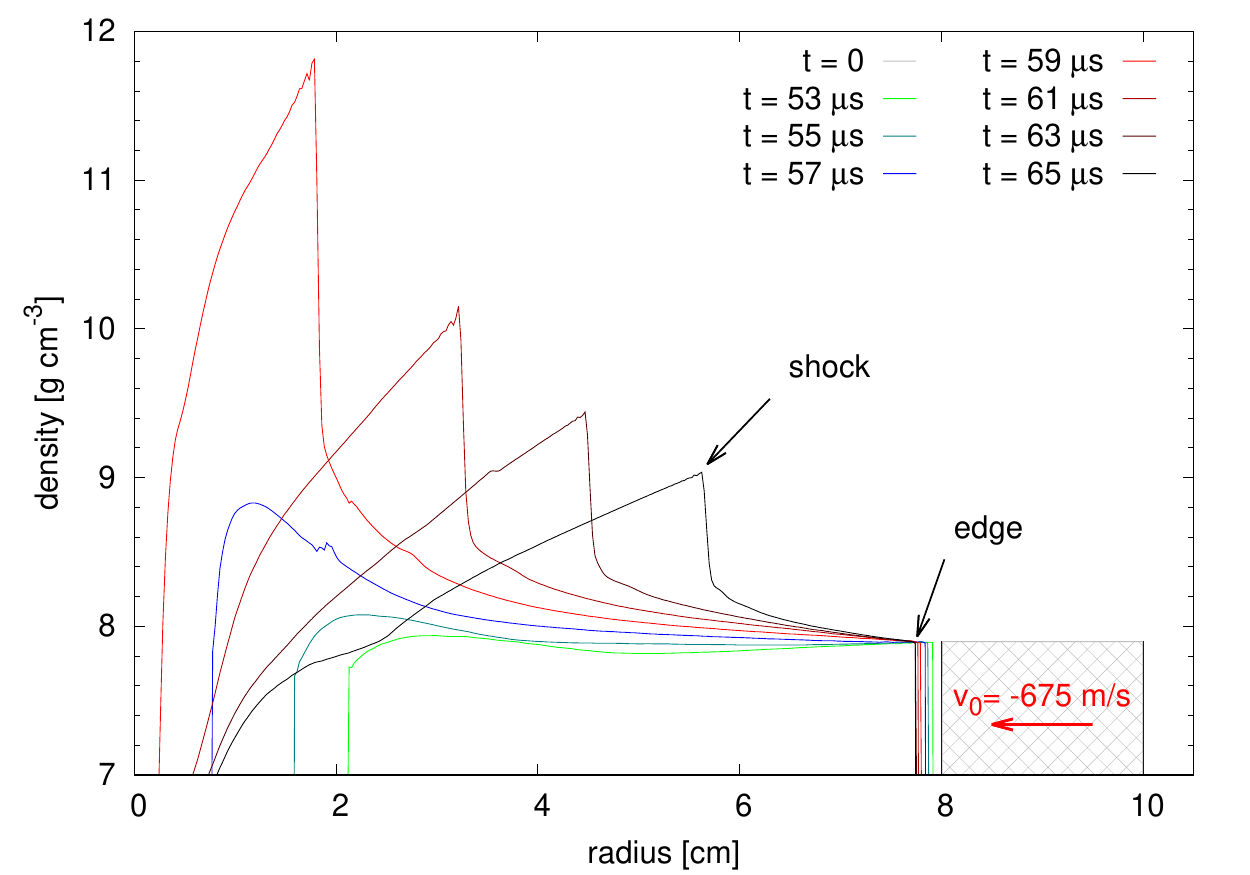}
  \caption{Evolution of the radial density profile in a representative case.
  Initially, the material (steel) is uniformly distributed in a spherical shell, 
  which is given a negative constant radial velocity, to initiate an implosion.
  Once the inner edge bounces, an expanding shock is formed propagating in the
  non-constant dynamic density background. We concentrate on the discontinuous 
  features of the post-bounce solution, namely shock and outer edge 
  (marked by arrows).} 
  \label{fig:implosion}
\end{figure}

To test our new approach to dynamic radiography, we studied the propagation of shocks generated by an imploding steel shell.
We limited our study to 1D spherical symmetry, a Mie-Gr\"uneisen (MG) equation of state (EOS), and a Steinberg-Guinan strength model applicable to metals at high strain rates \cite{steinberg1980constitutive}.
To simulate implosion dynamics, we ran the CTH hydrodynamic code \cite{hertel98a} on a uniform grid with 650 cells extending from the origin to a maximum radius of ${R_{D} = 16\ \cm}$, giving a grid cell size ${\Delta x = 246.15\ \um}$. Initial conditions for the shell were chosen as follows: uniform density equal to that of steel under standard atmospheric conditions,
inner and outer radii of ${R_{\rm in} = 8\,\cm}$ and ${R_{\rm out} = 10\;\cm}$, respectively, and a uniform implosion velocity of ${v_{\rm impl} = -675\ \mps}.$ (See Figure~\ref{fig:implosion}.) 

For each simulation, we generated synthetic radiographs at a series of times between 63 $\mu s$ -- 64.5 $\mu s$ in 0.5 $\mu s$ intervals using the Monte Carlo particle transport simulation software MCNP6 \cite{werner_mcnp6.2_2018}. From these synthetic images, we extracted shock and outer edge locations using standard edge-detection methods.  (see Figure~\ref{fig:density_direct_2D}.) The object center was placed 133 $cm$ from the source and the detector was placed 50 $cm$ from the object, resulting in a magnification of 1.376.  We irradiated the object with a 4 MeV mono-energetic X-ray source and generated a synthetic radiographic image using the radiography tally.  This is a quasi-deterministic calculation that maps each particle to each pixel.  Radiographs of direct and scattered radiation were generated separately and 1e3 and 5e5 particles were used, respectively.  Additionally, the scatter simulations employed variance reduction by using the PDS card, which calculates the probability of all possible interactions and significantly reduces rare event fliers from Compton scattering. \cite{werner_mcnp6.2_2018}  Experimental contributions that degrade the image, such as source blur, energy-dependent detector efficiency, and noise, were not included in these simulations.

We generated a dataset of simulations by varying parameters characterizing the MG equation of state~\cite{hertel98b}, 
\begin{align}
    P\left(\chi:=1-\frac{\rho_0}{\rho}, T\right) 
    = \frac{\rho_0 c_s^2\chi\left(1 - \frac12\Gamma_0\chi\right)}
           {(1 - s_1\chi)^2}
    + \Gamma_0\rho_0 c_V (T - T_0),
    \label{eq:mie-grueneisen}
\end{align}
over a 5-dimensional (5D) cube.
Here $\rho_0$ and $T_0$ are the reference density and temperature, respectively, $c_s$ is the speed of sound, $\Gamma_0$ is the Gr\"uneisen parameter at the reference state, $s_1$ is the slope of the linear shock Hugoniot, and $c_V$ is the specific heat capacity at constant volume.
Of the six parameters in \eqref{eq:mie-grueneisen}, we kept the reference density $\rho_0$ fixed and varied the vector $\sigma=(T_0, c_s, s_1, \Gamma_0, c_V)$ around its nominal value listed in Table~\ref{tab:nominal}. Each component of $\sigma$ was varied in $2\%$ intervals within $\pm 10\%$ of the nominal value. Each component therefore assumed 11 discrete values, implying a total dataset comprising ${11^5\equiv161051}$ simulations.

\begin{table}[htbp]
  \centering
  \begin{tabular}{lcccccc} 
   \hline \hline
   Material & $\rho_0\;[\gcc]$ & $T_0$ [K] & $c_0$ [mm/$\mu$s] & $s_1$ & $\Gamma_0$ 
            & $c_V$ $[{\rm erg}\;{\rm g}^{-1}\;{\rm eV}^{-1}]$\\
   \hline
   Steel    & 7.896 & 293.15 & 4.569 & 1.49 & 2.17 & $5.18\times10^{10}$ \\
   \hline \hline
  \end{tabular}
  \caption{Nominal parameter values in the employed Mie-Gr\"uneisen equation of state.
           The reference density parameter $\rho_0$ is fixed.}
  \label{tab:nominal}
\end{table}

For testing purposes, we generated an 15 additional simulations, labeled J1-J15.
Simulations J1-J10 were performed using the MG equation of state, while J11-J15 used
a SESAME equation of state \cite{johnson1994sesame}.
Parameters of these simulations are listed in Table~\ref{tab:jennies}.
\begin{table}[htp]
\begin{center}
\begin{adjustbox}{width=0.9\textwidth}
\begin{tabular}{l|lllll l l|ccc}
\cline{1-6}
\cline{8-11}
 & \multicolumn{5}{c}{EOS: Mie-Gr\"uneisen} & & & 
   \multicolumn{3}{c}{EOS: Sesame} \\
 & $T_0$& $c_s$ & $s_1$& $\Gamma_0$ & $c_v$ & & & material & strength & scaling \\
 &      &       &      &            &       & & &          &  model   &  ratio  \\
\cline{1-6}
\cline{8-11}
J1 &  0.2  &  -6.1 &   6.0 &  -6.0 & -10.0 & & J11 & STEEL\_HP & Steinberg & $1$    \\
J2 &  0.2  &   5.1 &   6.0 &  -6.0 & -10.0 & & J12 & STEEL\_HP & Steinberg & $0.99$ \\
J3 &  0.2  &   5.1 &   5.4 &  -6.0 & -10.0 & & J13 & STEEL\_HP & Steinberg & $0.98$ \\
J4 &  0.2  &   5.1 &   5.4 &   5.1 & -10.0 & & J14 & STEEL\_HP & PTW 304LSS1 & $0.98$ \\
J5 &  0.2  &   5.1 &   5.4 &   5.1 & -31.2 & & J15 & STEEL\_HP & PTW 304LSS1 & $0.99$ \\ 
\cline{8-11} 
J6 &-11.7  &   5.1 &   5.4 &   5.1 & -31.2 \\
J7 &-11.7  & -16.8 &   5.4 &   5.1 & -31.2 \\
J8 &-11.7  & -16.8 &  18.8 &   5.1 & -31.2 \\
J9 &-11.7  & -16.8 &  18.8 & -13.4 & -31.2 \\
J10&-11.7  & -16.8 &  18.8 & -13.4 &   7.4 \\
\cline{1-6}
\end{tabular}
\end{adjustbox}
\end{center}
\caption{Test simulations J1-J15 used to evaluate the quality of density reconstruction
and parameter estimation. All quantities are in terms of percentile deviations from the 
nominal values (see Table~\ref{tab:nominal}).
Simulation database admits a $\pm10\%$ variation in every parameter, so only the first
four simulations (J1-J4) are within the bounds of the database. The last 5 test 
simulations (J11-J15) are performed with a different EOS.Finally, cases (J14-J15) also use a Preston Tonks Wallace material strength model. \cite{preston2003model}  
} 
\label{tab:jennies}
\end{table}

\section{Reconstruction Using Hydrodynamic Features and Learned Dynamics: Data-Driven Modeling}
\label{sec:data-driven}
As discussed in the introduction, we are interested in reconstructing
density fields $\rho$ using only robust features of radiographs, which in
the current setting consist of the shock and edge locations.
As such, we formulate our learning problem 
as one of learning a map from sequences of shock and edge locations $F$ to a full density field $\rho$
and denote this process as $F\mapsto \rho$.
For training data we use a set of density field sequences obtained from our simulation database
and the corresponding shock and edge locations for each sequence. 

A challenging aspect of these data is that the mapping from features to density can be one-to-many.

Our approach to handling this aspect of the problem
is to assume that the underlying
(unknown) model generating the data is probabilistic.
Thus, instead of learning a map
$F\mapsto \rho$, we are led to learn to sample from the conditional probability $p(\rho\mid F)$ of
encountering a density field $\rho$ given a robust feature sequence $F$. 
In this setting, a (conditional; see later) generative model is
called for. This is because at its core a generative model includes
a representation of the {\em probability distribution} of the data,
e.g., as indicated above%

  While there are a handful of generative model
classes ranging from generative adversarial networks (GANs)
\cite{goodfellow2014generative} to
variational autoencoders  \cite{kingma2013auto} to continuous
normalizing flows \cite{rezende2015variational} to diffusion
maps \cite{coifman2006diffusion} and others, we focus attention on GANs given the extreme level of
activity seen in this class of models since their introduction in
2014.  GANs are a powerful class of generative models that are trained using an adversarial technique:
while a discriminator continually learns to discriminate
between the generator’s synthetic output and true data, given a noise
source, the generator progressively seeks to fool the discriminator by
synthesizing yet more realistic samples. The process continues until a
quasi-equilibrium is reached, wherein the discriminator's feedback to
the generator does not permit further large improvements of the
generator, but rather leads to stochastic variations that explore the
local basin of attraction.

Initially, the training of GANs was problematic in that they commonly
suffered from instability and various modes of failure. One of the
approaches developed in response to the problem of instability in the
training of GANs was the Wasserstein GAN (WGAN). \cite{arjovsky2017towards} The WGAN approach
leverages the Wasserstein distance \cite{kantorovitch1958translocation} 
on the space of probability distributions to produce a value function that
has better theoretical properties, and requires the discriminator
(called the critic in the WGAN setting) to lie in the space of
1-Lipschitz functions. While the WGAN approach made significant
progress towards stable training of GANs and gained popularity,
subsequent accumulating evidence revealed that the WGAN approach could
sometimes generate only poor samples or, worse, even fail to
converge.
In further response to this aspect of WGANs, a variation of
the WGAN was developed: while in the original WGAN, a {\em weight-clipping} 
approach was used to ensure that the critic lay in
the space of 1-Lipschitz functions, the new approach used a {\em gradient
penalty} term in the critic's loss function to satisfy the 1-Lipschitz
requirement. As such this approach was called the Wasserstein
GAN with gradient penalty (WGAN-GP).

Indeed, in the preceding part of this section, when we have used the
terminology of generative models and GANs, we have really been talking
about the conditional form of the generative model(s) since that was
natural for the problem on hand. We note that if indeed we are
interested just in drawing new realistic samples of the density field
given the set of numerical simulated density fields, without further
requirement on the shock and edge locations (to be as given by the
radiographs), then the above procedure reverts to an unsupervised
learning task, and the generative model seeks to learn simply the
distribution of the density field data $p(\rho)$ rather than the conditional distribution $p(\rho\mid  F)$ as previously introduced. 

In the following subsections we explore two methods of
generating a sequence of density fields conditioned on the shock and edge locations.
In the first approach, we use a conditional WGAN (cWGAN) model,
trained with shock and edge locations extracted with pixel level resolution.
In the second approach, we use the subpixel feature extraction method (detailed in Appendix~\ref{sec:subpixel}) to condition a cGAN. 
Similar to \cite{isola2017image}, 
we incorporate an $L^1$ reconstruction loss term for the generator update, 
as well as additional discrete modules to enforce 
prior knowledge about the density reconstruction 
such as the known mass and feature locations.
The examination of two different conditional GAN architectures,  resolutions at which the features may be extracted from the radiographs, as well as inclusion of physics based constraints in one network architecture allows for a more thorough investigation of the properties of the solution space and behavior of our mapping from feature space to density fields.

\subsection{Conditional Wasserstein GAN (cWGAN)}
\label{sec:wgan}
As a first approach,
we train a cWGAN conditioned on 
shock and edge locations.
As a first approach,
we train a cWGAN conditioned on 
shock and edge locations.
The discriminator loss function is
\begin{align}
  \mathop{\mathbb{E}}_{\rho \sim P, z \sim \mathcal{N}} 
  [\disc(\gen(z, F(\rho)), F(\rho))]
  -
  \mathop{\mathbb{E}}_{\rho \sim P} [\disc(\rho, F(\rho))] 
  +
  \lambda \mathop{\mathbb{E}}_{\rho \sim P_u} \left[\left(\lVert\nabla_\rho
  \disc(\rho)\rVert_2 - 1\right)^2\right].
  \label{eq:wgan-dloss}
\end{align}
Here, 
 $\rho \sim P$
corresponds to real samples
(that are numerically generated using the
hydrodynamic solver and which are used for training the cWGAN)
drawn uniformly at random from the training set, 
$\disc$ refers to the discriminator network with parameters $\discparams$,
$\gen$ refers to the generator network with parameters $\genparams$,
$z \sim \mathcal{N}(0,1)$ is i.i.d. Gaussian noise,
$\lambda$ is a coefficient that corresponds to the
penalty,
and the final term in the sum is the gradient penalty
that encourages the discriminator to be 1-Lipschitz;
see \cite{gulrajani2017improved} for further details.

On our experiments,
the generator and
discriminator were multi-layer perceptrons (MLP) with two hidden
layers, each consisting of 512 neurons; a ReLU activation was used and
a dropout of 10\% of the nodes was implemented to discourage overfitting.

\subsection{Conditional GAN with Data Fidelity Constraints (cGAN-DF)}
\label{sec:cgan}

\begin{figure}
  \centering
  \begin{tabular}{cc}
     Discriminator & Generator
     \\
     \includegraphics[width=.4\textwidth]{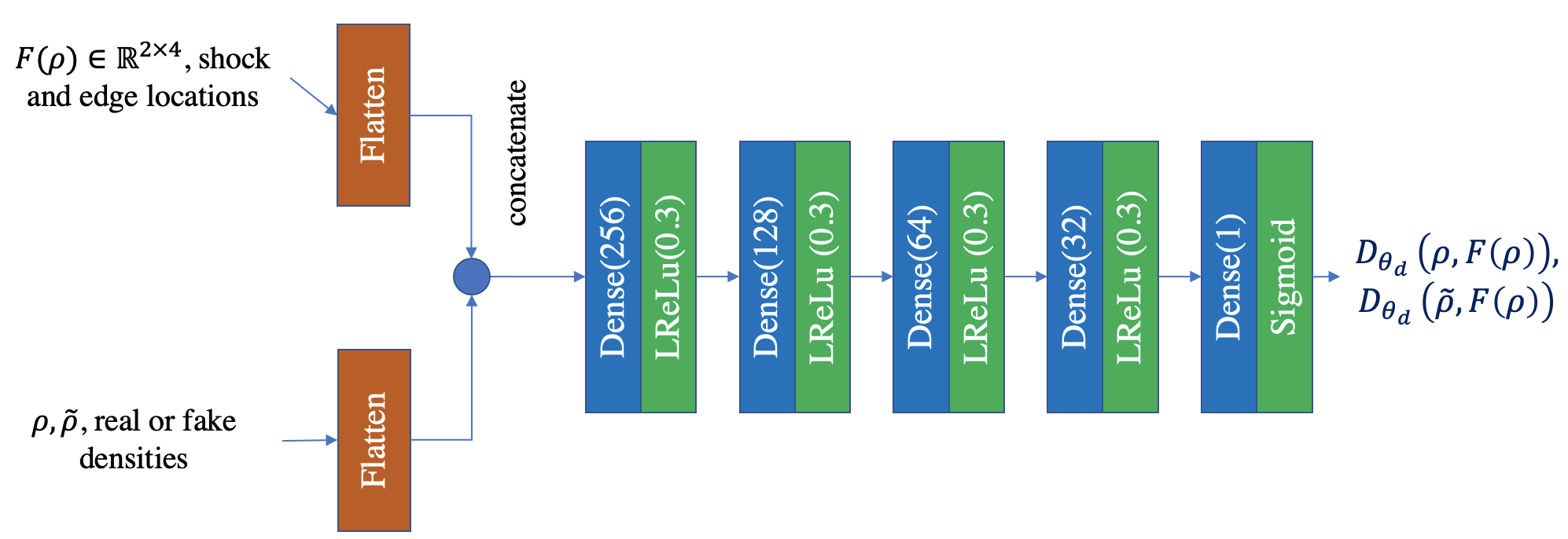} &
     \includegraphics[width=.4\textwidth]{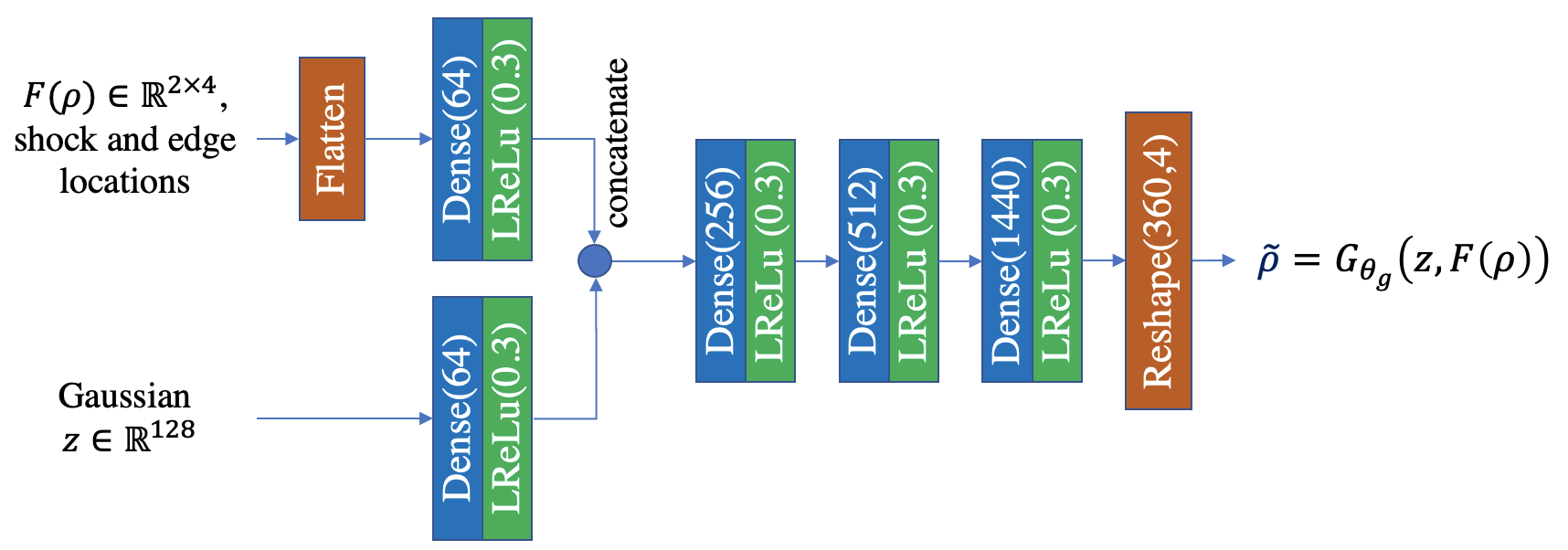}
  \end{tabular}
  \caption{Discriminator and generator architectures for cGAN-DF.
  } 
  \label{fig:cgan_arch}
\end{figure}

Unlike the cWGAN case, in the traditional GAN setting further modifications of the generator loss function were required to obtain meaningful results and these are described below.
In this second approach, we build on a traditional cGAN by incorporating modules 
that enforce data fidelity constraints.
Figure~\ref{fig:cgan_arch} presents the cGAN-DF architecture.

The discriminator loss function is 
\begin{equation}
    -\underset{\rho \sim P}{\mathbb{E}}
    \left[\log(\disc(\rho,F(\rho))) \right]
    - \underset{\rho \sim P,z\sim \mathcal N}{\mathbb{E}}
    \left[\log(1-\disc(\gen(z,F(\rho)),F(\rho)))\right]
\label{eq:cgan_dis_loss}
\end{equation}

\noindent where 
$\rho \sim P$ denotes a density drawn uniformly at random from the training set,
$z \in \mathbb R^{128}$ is i.i.d. Gaussian noise, 
i.e., $z \sim \mathcal N(\mathbf 0, \mathbf I_{128})$.
$F(\rho) \in \mathbb R^{2\times 4}$ are instances of shock and edge locations on which the cGAN-DF is conditioned, 
and $\rho \in \mathbb R^{360\times 4}$ are instances of 1D density profiles produced via simulation.
Likewise,  $\tilde \rho = G_{\theta_g}(z,F(\rho))$ are generated (also called fake) densities.

In a traditional cGAN, the loss function for the generator is 
\begin{equation}
L_G(\theta_g)=
    \underset{\rho \sim P, z\sim \mathcal N}{\mathbb{E}} \left[\log(1-D_{\theta_d}(G_{\theta_g}(z,F(\rho)),F(\rho)))\right].
\label{eq:cgan_gen_loss}
\end{equation}

\noindent 
We augment this loss function with an $L1$ reconstruction loss term as in \cite{isola2017image}
\begin{equation}
 L_{L1}(\theta_g) = 
\underset{\rho \sim P,z\sim \mathcal N}{\mathbb{E}} \left[\left\Vert  
\rho - \gen(z,F(\rho)) \right\Vert_1\right]
    \label{eq:cgan_l1_loss}
\end{equation}

Prior to training the cGAN-DF, density simulations from the training set are processed 
using the method described in Appendix~\ref{sec:subpixel} to produce subpixel shock and edge locations.
We denote this mapping from density to features $F(\rho)$. 
Ideally, generated densities show backwards consistency with the features, i.e., $F(\gen(z,F(\rho)))=F(\rho)$.
To enforce this constraint, we introduce an $L2$ penalty on differences in $F(\rho)$ and $F(\gen(z,F(\rho)))$.
However, the edge and shock extraction procedure, $F$, is not differentiable and cannot facilitate back propagation during GAN training.
Instead, a CNN model, denoted $\tilde{F}$ is pretrained on $(\rho,F(\rho))$ pairs to approximate this map.
$\tilde{F}$ is made up of five Conv1D layers of widths from 16 up to 256 with kernels of length three and LReLU(0.3) activations. 
The final layer is a Dense operation with a TanH activation and reshaping. 
The weights of $\tilde F$ are held fixed while training of cGAN-DF.
The corresponding loss term is given by
\begin{equation}
    L_{\tilde F}(\theta_g) = 
    \underset{\rho \sim P,z\sim \mathcal N}{\mathbb{E}} \left[
    \left\Vert F(\rho) - \tilde F \left(G_{\theta_g}(z,F(\rho))\right) \right\Vert_2
    \right]
    \label{eq:cgan_r_loss}
\end{equation}

Let $M(\tilde \rho) \in \mathbb R^4$ be the spherical integral of 
the radial density profile for the four distinct timeframes. 
We introduce an $L2$ penalty between the mass of the generated result
and known mass of the steel capsule, $M_\rho$.
\begin{equation}
    L_{M}(\theta_g) = 
    \underset{\rho \sim P,z\sim \mathcal N}{\mathbb{E}} \left[
    \left\Vert M_\rho - M\left(G_{\theta_g}(z,F(\rho))\right) \right\Vert_2
    \right]
\end{equation}

The resulting composite loss function for the generator update is 
\begin{equation}
  \mathcal{L}_{cGAN} (\theta_g)
  = \lambda_D L_G(\theta_g)
  + \lambda_{L1} L_{L1}(\theta_g)
  + \lambda_{\tilde F} L_{\tilde F}(\theta_g)
  + \lambda_M L_{M}(\theta_g)
\label{eq:GANcomp_loss}
\end{equation}

\section{Results}
\label{sec:results}
In this section, we describe and compare the results obtained from the two different network architectures.

\subsection{cWGAN Results}
\begin{figure}[htbp]
  \centering
  \begin{tabular}{cc}
   \includegraphics[width=0.41\textwidth]{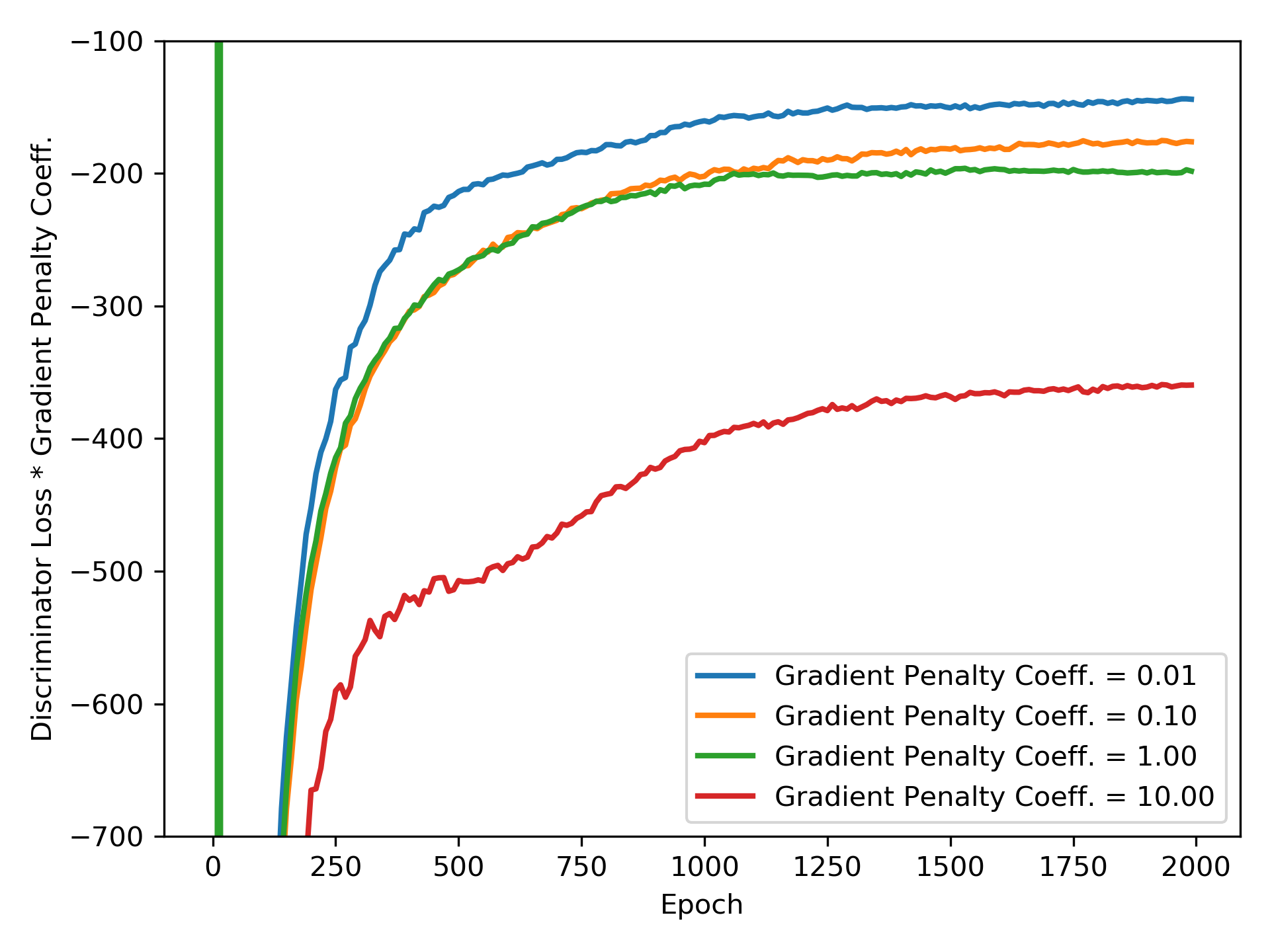}&
   \includegraphics[width=0.47\textwidth]{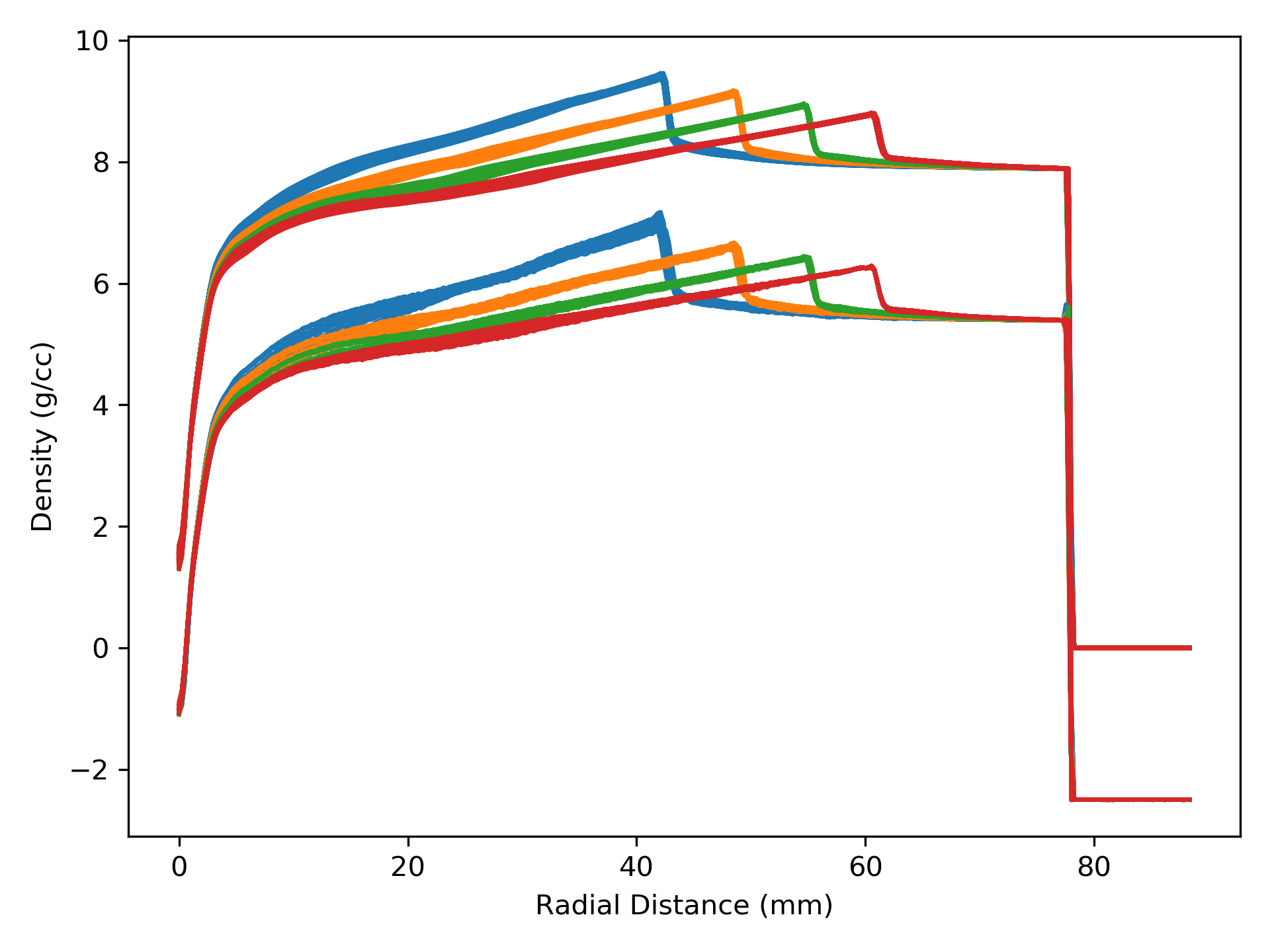} 
  \end{tabular}
  \caption{The left panel shows the discriminator loss as a function
    of training epoch. The loss is averaged over ten epochs to reduce
    the number of data points. The averaging also leads to smoother
    loss curves.
    When the coefficient associated with
    the gradient penalty $\lambda$ in (\ref{eq:wgan-dloss}) is varied
    over three orders of magnitude ($0.01\ge\lambda\ge 10$) the cWGAN
    trains reliably and stably. The right panel verifies the ability
    of the trained generator, given a sequence of shock and edge
    locations, to reconstruct a range of density profiles (bottom
    ensemble of density profiles) that is commensurate with the range
    of density profiles seen in the reference computations (top
    ensemble of density profiles). The cWGAN reconstructed ensemble is
    offset by -2.5 g/cc to improve clarity.}
\label{fg:wgan1}
\end{figure}

The left panel of Fig.~\ref{fg:wgan1} shows the discriminator loss as
a function of training epoch.  Training can be seen as consisting of
three phases: an initial phase (up to about 100 epochs) when there are
large changes given the random initial condition resulting both in a
poor generator and the discriminator not knowing the Wasserstein
metric. On incremental learning on the parts of both the generator and
the discriminator, the initial phase of violent changes is followed a
period of slower changes (between epochs 100 and 1000). In the final
phase, there is an asymptotic approach to equilibrium characterized by
extended periods of stable behavior of both the generator and
discriminator. In this figure, it is further observed that when the
coefficient associated with the gradient penalty $\lambda$ in
(\ref{eq:wgan-dloss}) is varied over three orders of magnitude
($0.01\le\lambda\le 10$) the cWGAN trains reliably and stably without
suffering any of the modes of failure such as mode-collapse that are
common in the training of GANs. In the following, we show results for
the smallest value of the gradient-penalty coefficient (0.01) while
noting that the results for the other values of the gradient-penalty
coefficient differ so slightly as to be visually indistinguishable
most often for the diagnostics considered. This further confirms the
robustness of the WGAN results presented and discussed.

The top set of density fields in the right panel of
Fig.~\ref{fg:wgan1} correspond to the most frequently occurring
time-sequence of shock and edge locations in the numerical
simulations. When this sequence of shock and edge locations are
input to the trained generator (of the cWGAN) along with a further random
number,
one set of four density profiles corresponding to the four
shock and edge locations is obtained. The lower set of density fields
show an ensemble of one hundred such predictions. This constitutes an
initial verification of the ability of using a cWGAN to reconstruct
a full density profile given just the shock and edge locations.

Next, we further examine the ability of the trained generator
to reconstruct the density field given
shock and edge locations where now the shock and edge locations come
from test numerical computations, some of which use an
equation of state that is significantly different from that used to
generate the data that was used for training the cWGAN. We consider
cases J1, J10, J11, and J15 in Table~\ref{tab:jennies}. Furthermore, since we have the
density profiles for these cases, we additionally examine the ability
of the trained discriminator to establish similarity to (or likewise, tell them
apart from) the density profiles used for training.

\begin{figure}[htbp]
  \centering
  \begin{tabular}{cc}
   \includegraphics[width=0.47\textwidth]{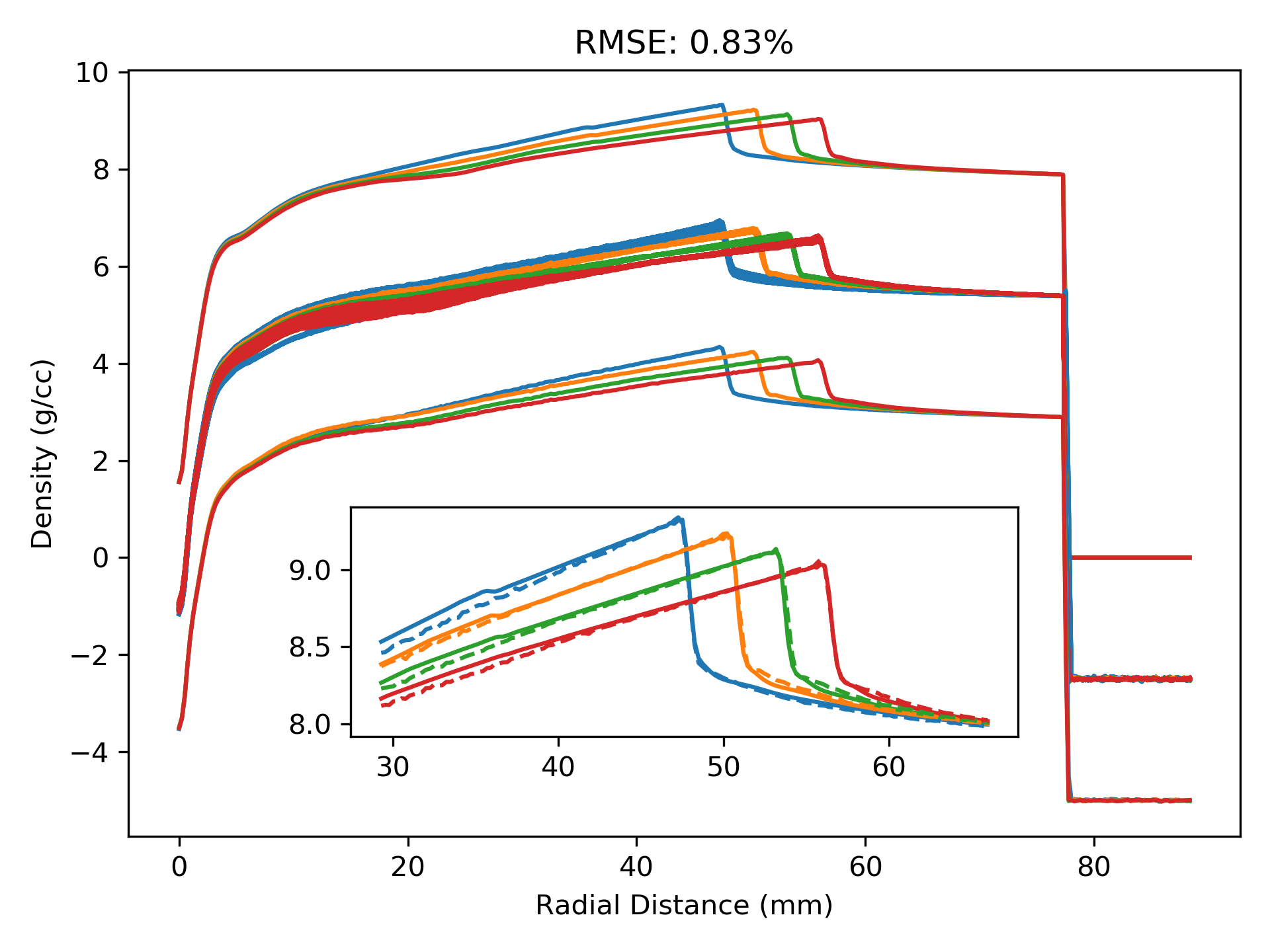} &
   \includegraphics[width=0.47\textwidth]{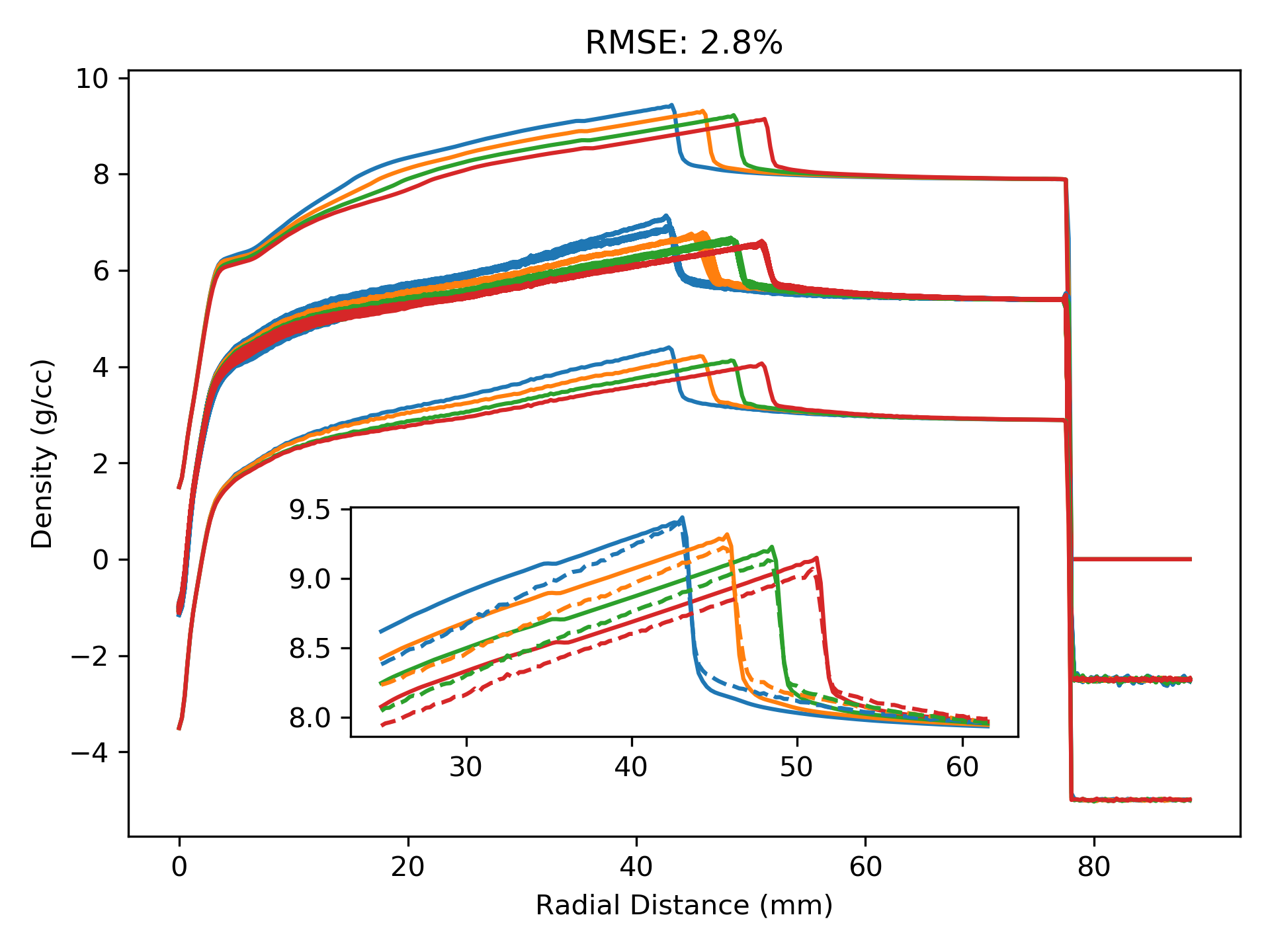}
   \\
   \includegraphics[width=0.47\textwidth]{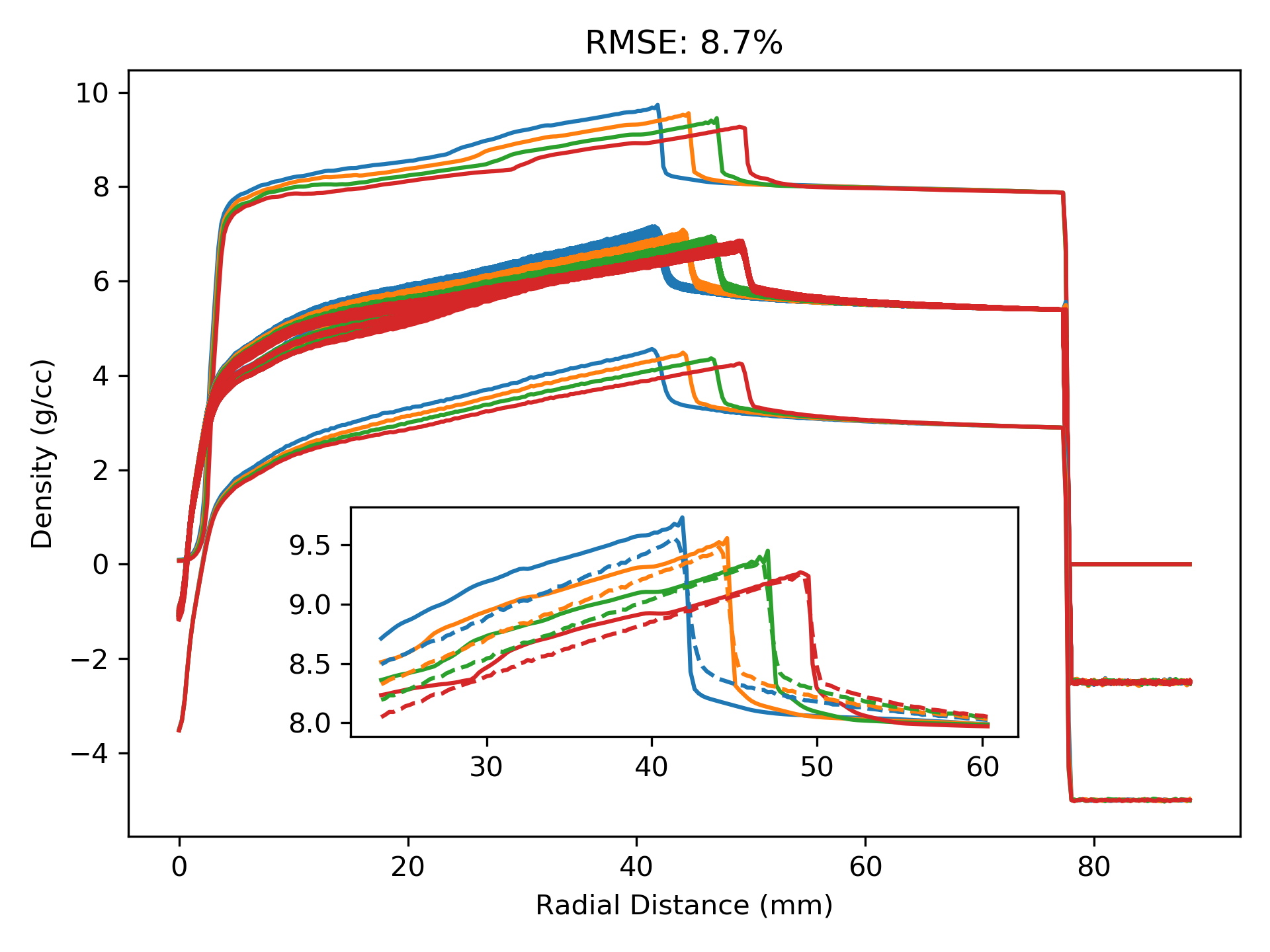} &
   \includegraphics[width=0.47\textwidth]{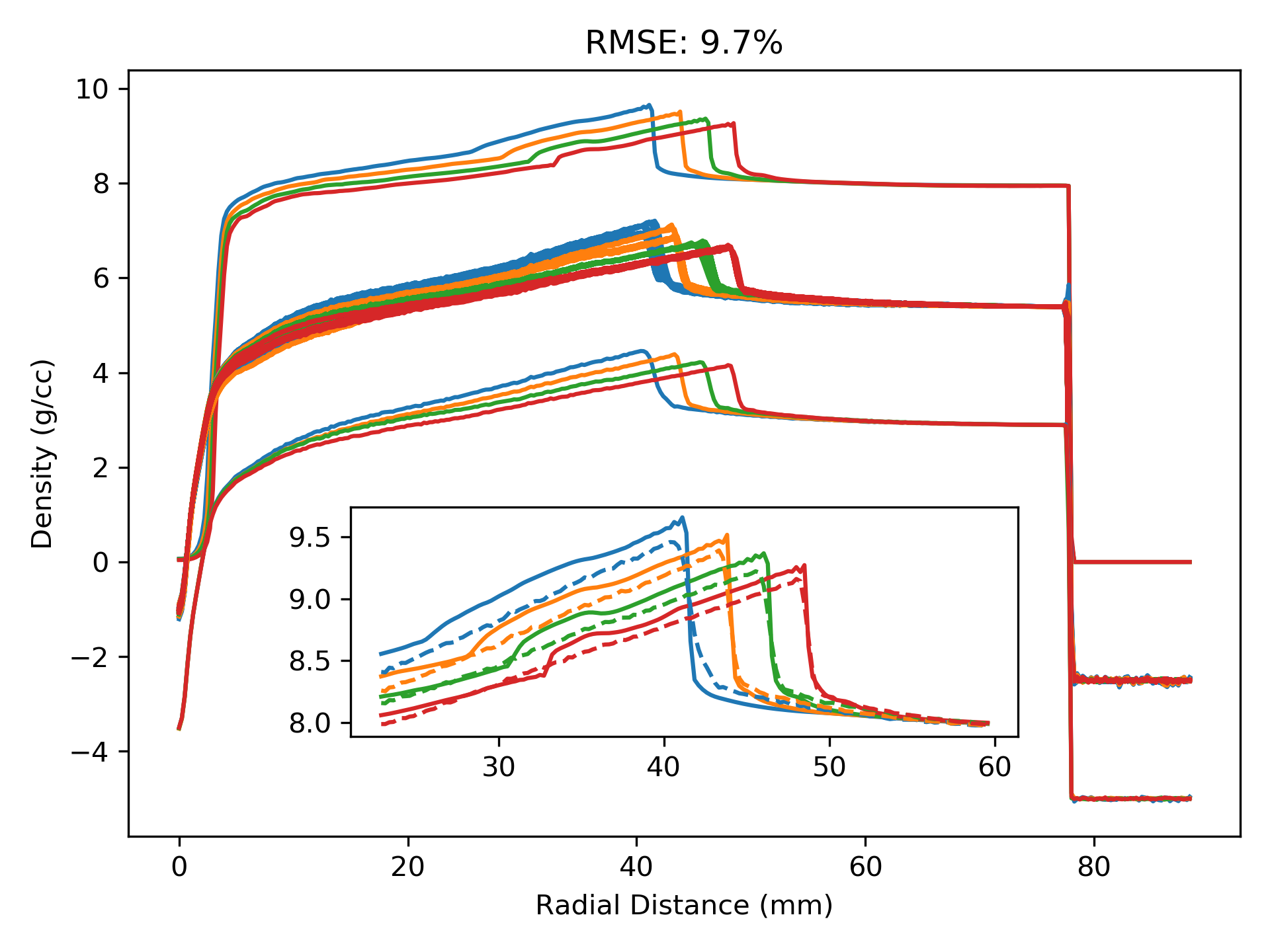}
  \end{tabular}
  \caption{Top: Cases 1 (left) and 10 (right); Bottom: Cases 11 and
    15. In each of these figures, the top set of density fields come
    from the test numerical integrations and form the reference to
    judge the quality of the cWGAN predictions. The next set
    of density fields comprise an ensemble of a thousand predictions
    from the cWGAN; they are offset by -2.5 (g/cc) to facilitate
    comparison. The final set of density fields that are offset by a
    further $-2.5\;\gcc$ correspond to the mean over the thousand cWGAN
    predictions. The inset is a zoom of the reference (solid)
    and predicted (dashed) density fields at the 
    four times considered. The RMSE is indicated on
    top. Larger differences between the equations of state used in
    cases 11 and 15, as compared to the equations of state used for
    the training data lead to larger density reconstruction errors in
    those cases.}
  \label{fg:wgan2}
\end{figure}

Figure~\ref{fg:wgan2} compares the cWGAN reconstruction of the density
profiles from the shock and edge locations for cases J1, J10, J11 and J15.
Each panel in the Figure additionally indicates the root mean 
square error (RMSE) between the average of the ensemble of generated 
density profiles and the ground truth 
(see also Table~\ref{tab:params-estimation}).

The RMSE of the cWGAN density reconstruction in case J1 is small
and the predicted profile is seen to match the reference numerical
profile well. However, the RMSE is seen to be larger for cases J11
and J15 with corresponding large mismatches in the density
behind the shock. Indeed, it turns out that the equation of state
used in case J1 is well represented in the training data whereas
that is not the case with cases J11 and J15.

\begin{figure}[htbp]
  \centering
  \includegraphics[width=0.7\textwidth]{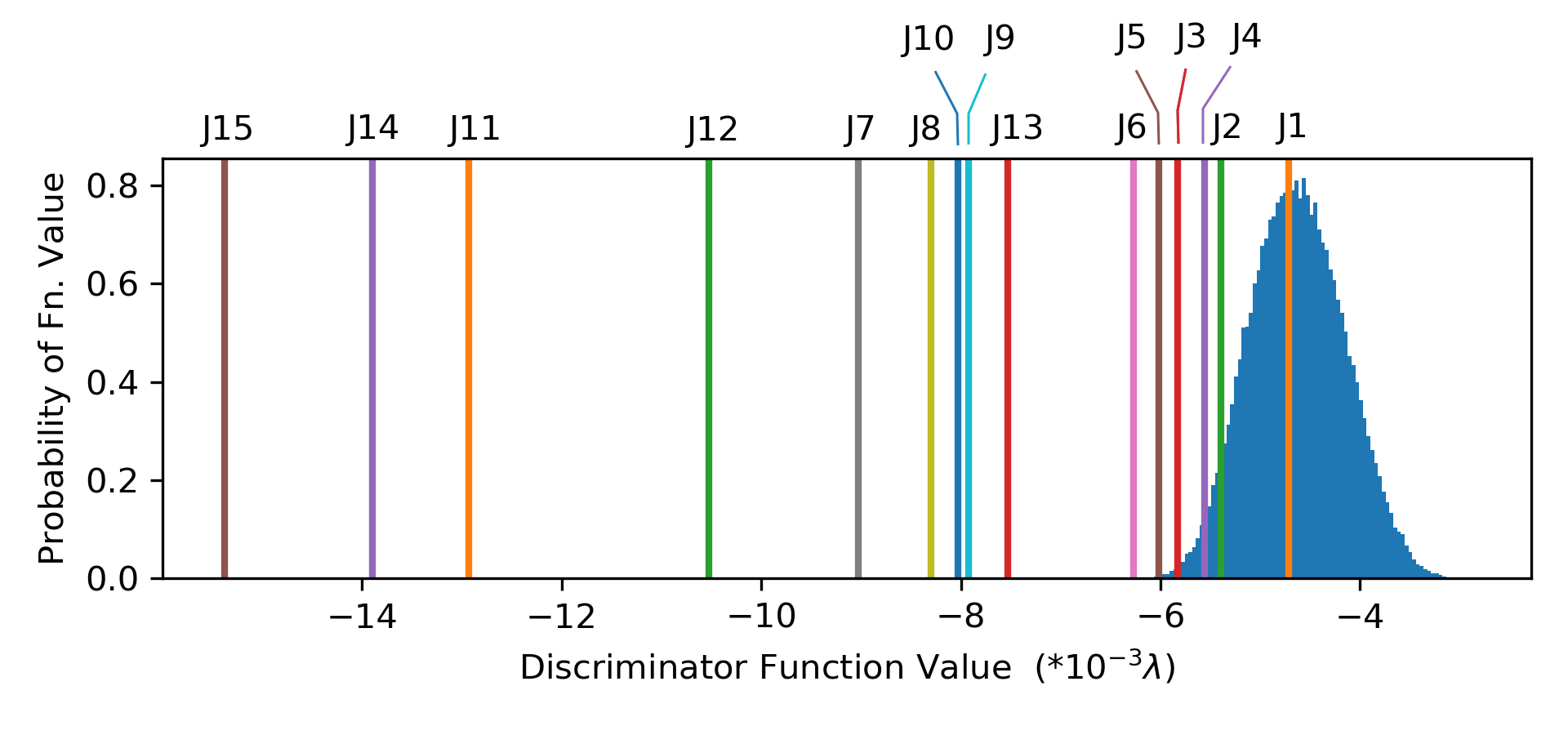}
\caption{Discriminator function values. Histogram shows the probability
  distribution in the training dataset. Delta functions for test cases 
  J1 through J15 are shown in different colors.}
\label{fg:wgan4}
\end{figure}

This is further borne out in terms of the discriminator suggesting
that while the density profile of case J1 is well within the kinds of density profiles
seen in the training data that is not the case for  J11 and
J15. This is can be seen in Fig.~\ref{fg:wgan4} which shows the
discriminator function value for the training set of density
profiles as a histogram and for each of the test cases J1 through
J15 as delta functions.

In the noisy environment of a hydrotest, scatter can affect the
recovery of even the most robust of features such as the shock and
edge locations. For example, in test case J1, experimentation with
the level of scatter suggests that the location of the shock can be, 
depending upon experimental design,
perturbed by as much as 300 microns whereas the location of the edge can
be perturbed by as much as 100 microns (while noting that the hydro
simulations here use a uniformly space grid with a mesh size of 246.15
microns).

\begin{figure}[htbp]
  \centering
  \includegraphics[width=0.7\textwidth]{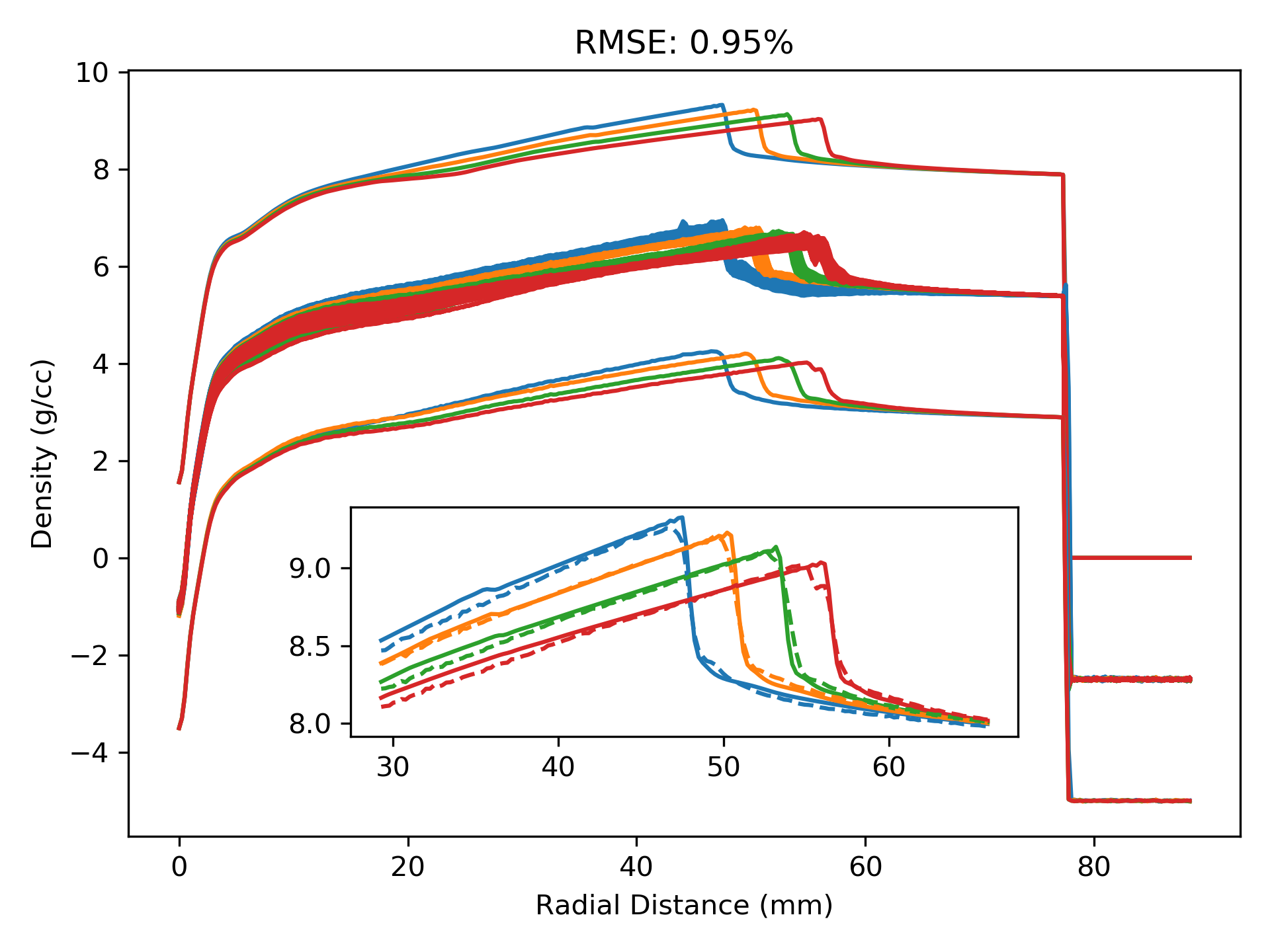}
\caption{Effects on the reconstruction of density of scatter-induced uncertainty in
the location of shock and edge. A shock location uncertainty of 300 microns and an
edge location uncertainty of 100 microns lead to significant changes in the reconstructed
density profiles. The density reconstruction error increases by a modest amount of
about 14\%.}
\label{fg:wgan-prtrb}
\end{figure}

Figure \ref{fg:wgan-prtrb} shows the results of reconstructing the
sequence of density profiles when the shock and edge locations are
perturbed as stated above, but when the reference profiles are left
unperturbed.  Again a thousand perturbations were considered. On
comparison with the top left panel of Fig. 10, changes in the location
of the shock of individual profiles is evident; because of the large
number of profiles plotted, this perturbation can only be seen as a
thickening of the envelope of profiles at the location of the
shocks. The result of the uncertainty of the edge location can also be
seen, but is significantly smaller, and this is commensurate with the
lower level of uncertainty in locating the edge due to scatter from
radiographs. Small changes in the structure of the shock near the
downstream end can also be seen at the different times of the sequence
of ensemble-averaged profiles in the zoomed inset. Finally, on
quantifying the effect of uncertainty in the shock and edge location
due to scatter, we see the overall RMSE increases by a modest amount
of about 14\% assuming the stated uncertainties in the shock and edge
positions.

\subsection{cGAN-DF results}

\begin{figure}
\centering
\footnotesize
\begin{tabular}{ccc}
  \includegraphics[width=0.3\textwidth]{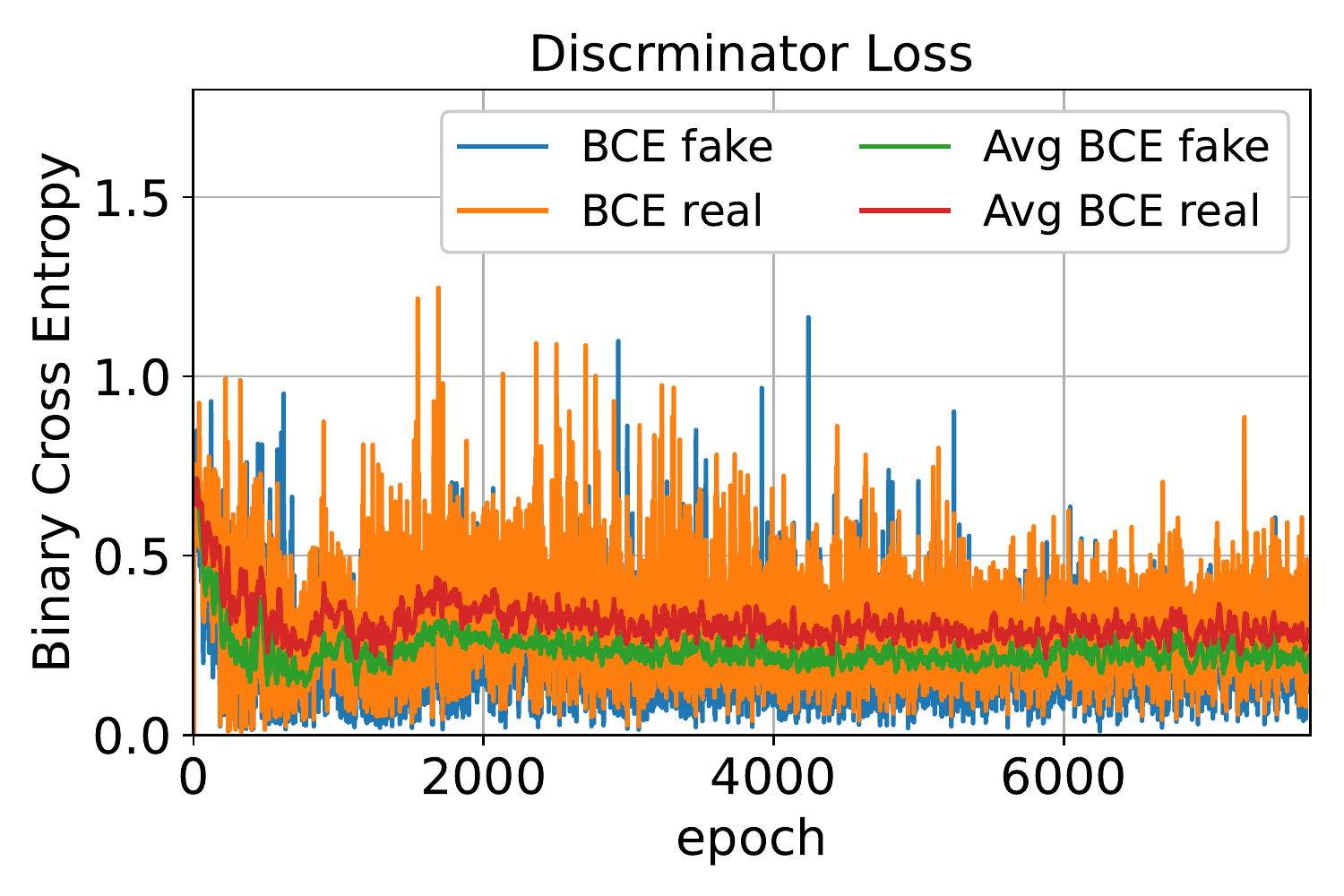} &
  \includegraphics[width=0.3\textwidth]{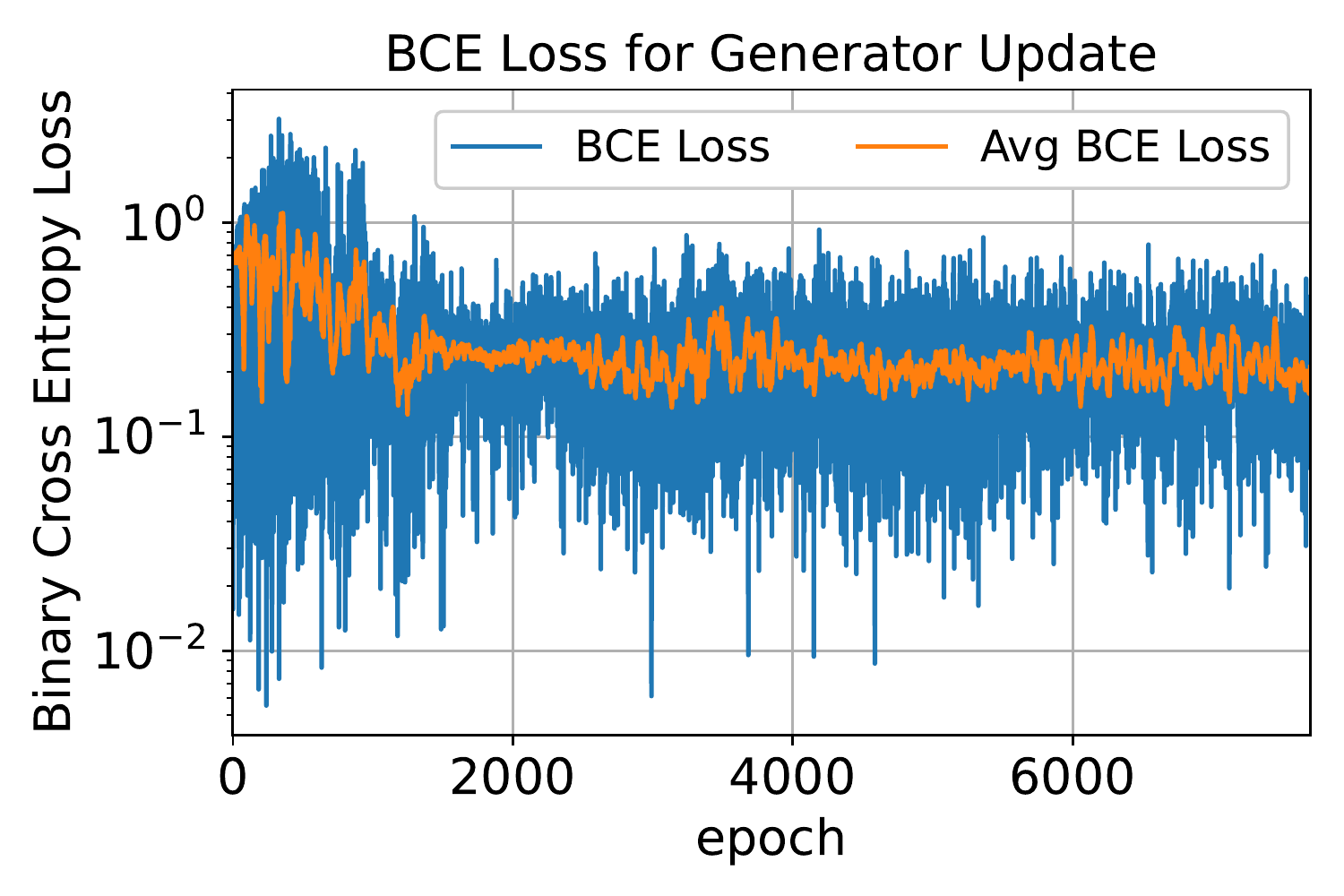} &
  \includegraphics[width=0.3\textwidth]{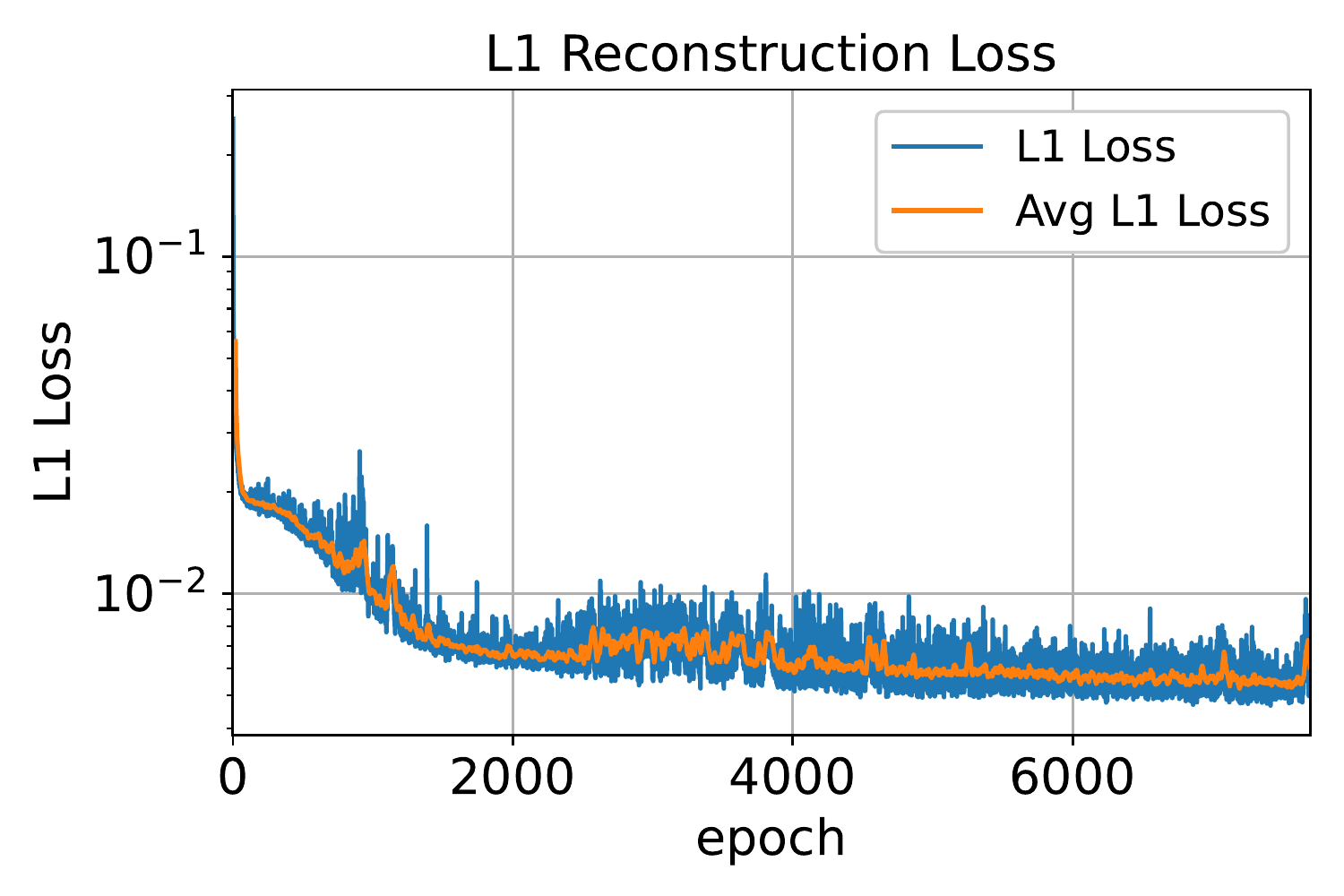}
  \\
  $L_{D}(\theta_d)$ &
  $L_{G}(\theta_g)$ &
  $L_{L1}(\theta_g)$
  \\
  \includegraphics[width=0.3\textwidth]{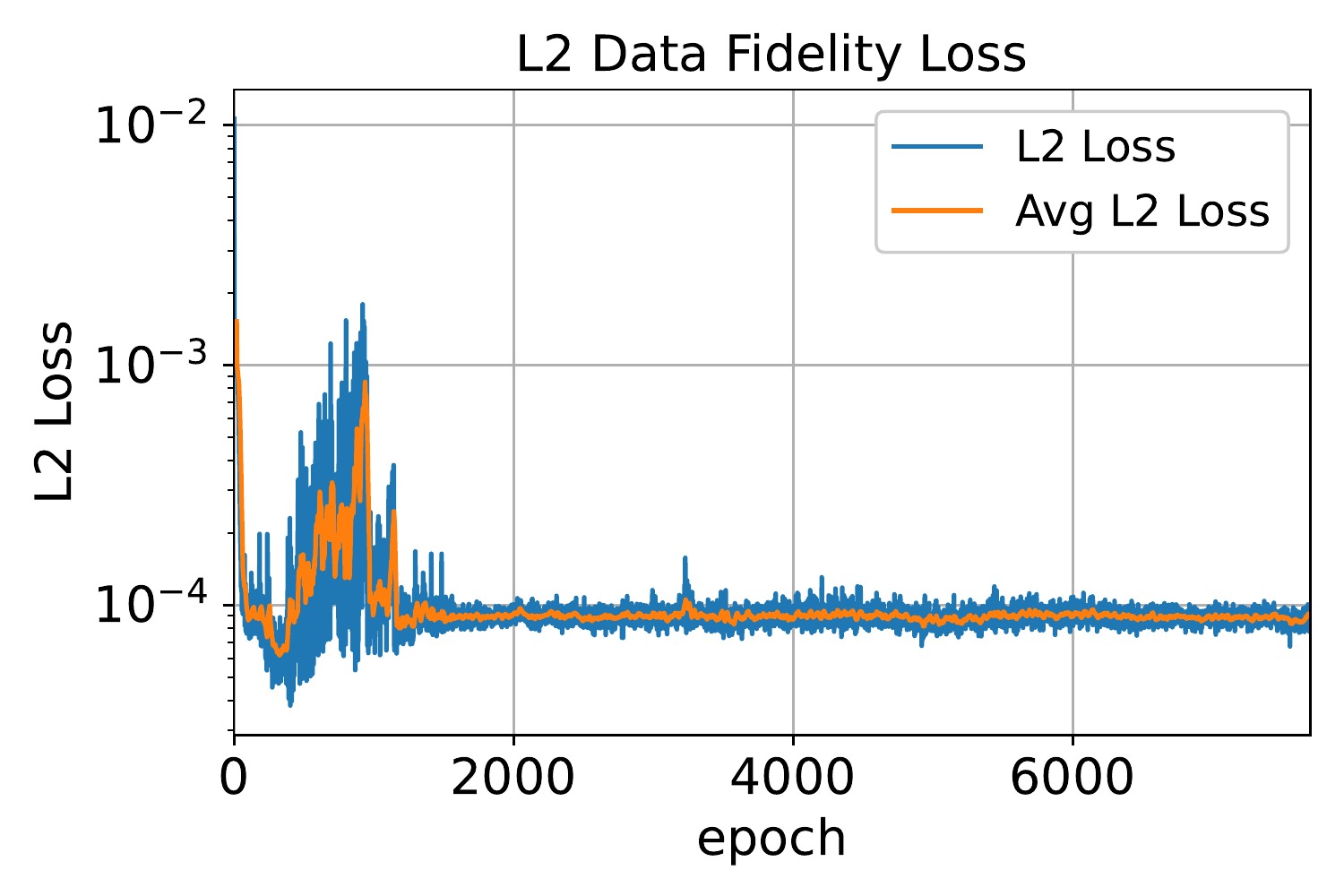} &
  \includegraphics[width=0.3\textwidth]{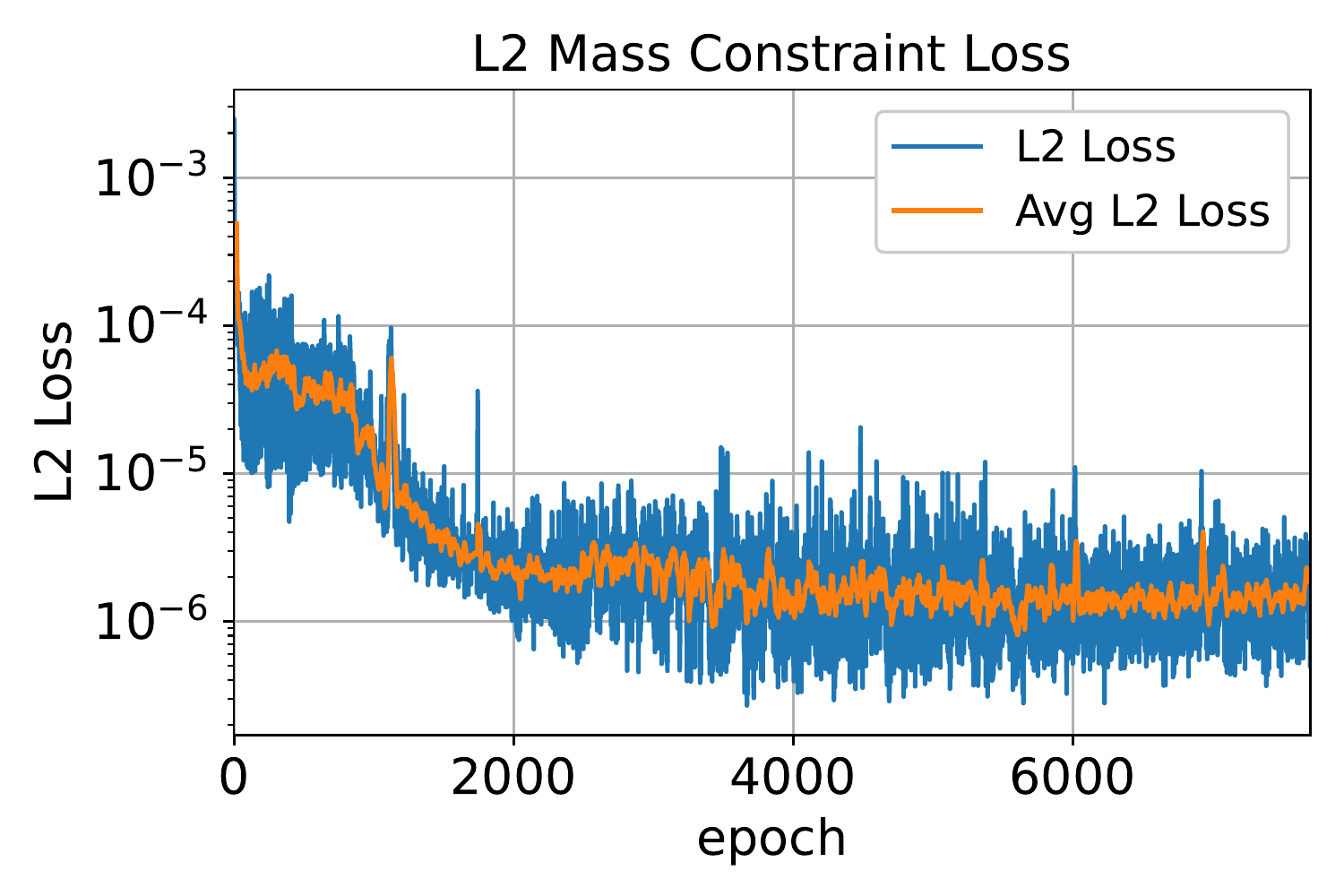} &
  \includegraphics[width=0.3\textwidth]{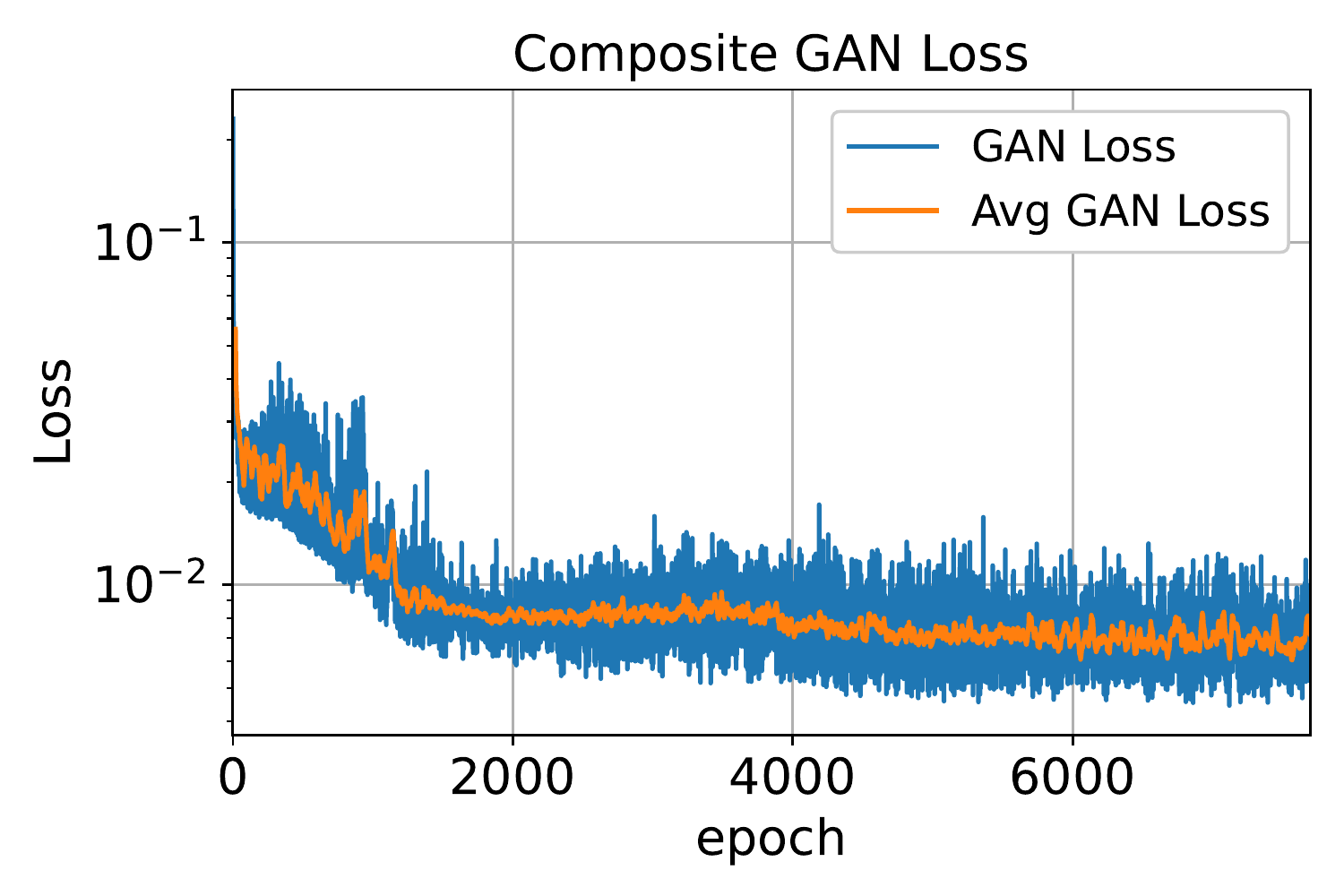}
  \\         
  $L_{\tilde R}(\theta_g)$ &
  $L_{M}(\theta_g)$ &
  ${L}_{cGAN} (\theta_g)$
\end{tabular}
\caption{cGAN-DF loss functions, plotted alongside running averages of 20 epochs for clarity.}
\label{fig:mh_cganloss}
\end{figure}

All models, including $\tilde R$, are optimized using Adam 
with a learning rate of $2\times10^{-4}$
and a $\beta_1$ parameter 0.5.
The cGAN-DF training is terminated when shock locations appear sharp, at 5000 epochs, to prevent overfitting due to the L1 loss term.
All results utilize, 
$\lambda_D=1$, $\lambda_{L1}=100$, $\lambda_{\tilde F}=10$, and $\lambda_M=1$.
The loss functions are shown in Fig.~\ref{fig:mh_cganloss}.

\begin{figure}
\centering
\footnotesize
\begin{tabular}{ccc}
  \includegraphics[width=0.45\textwidth]{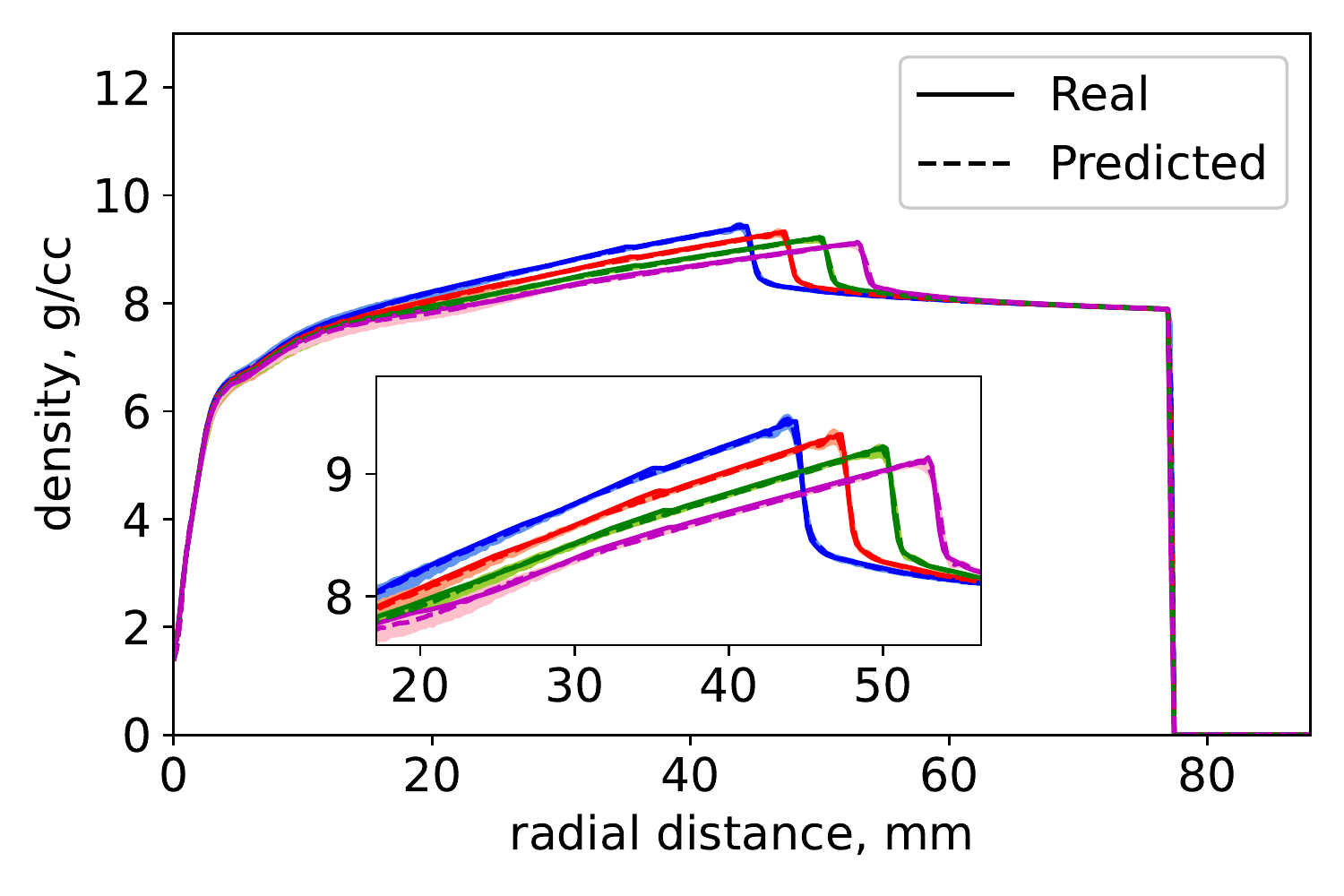} &
  \includegraphics[width=0.45\textwidth]{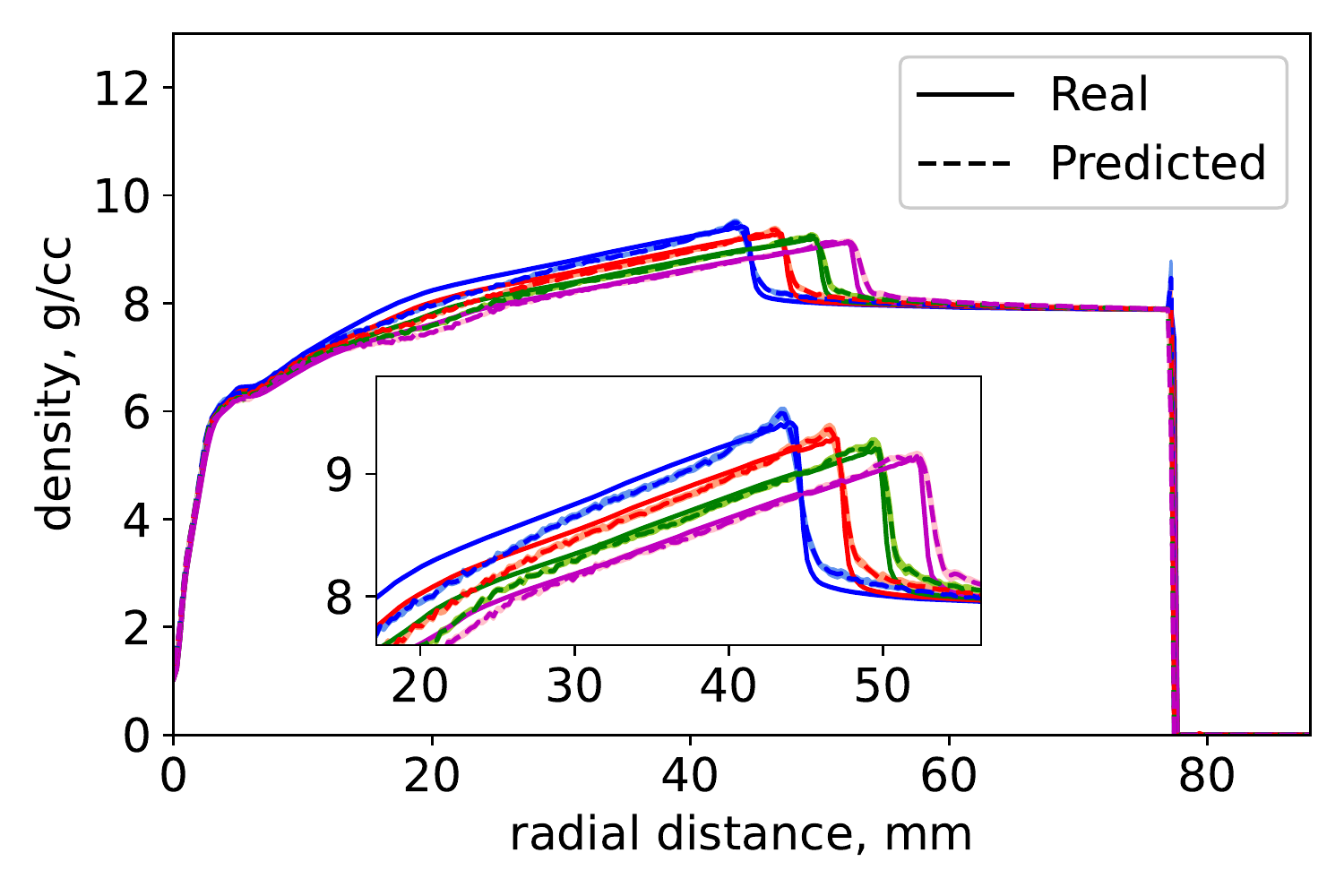}
  \\
  (a) RMSE $=0.48\%$ &
  (b) RMSE $=5.03\% $
  \\
  \includegraphics[width=0.45\textwidth]{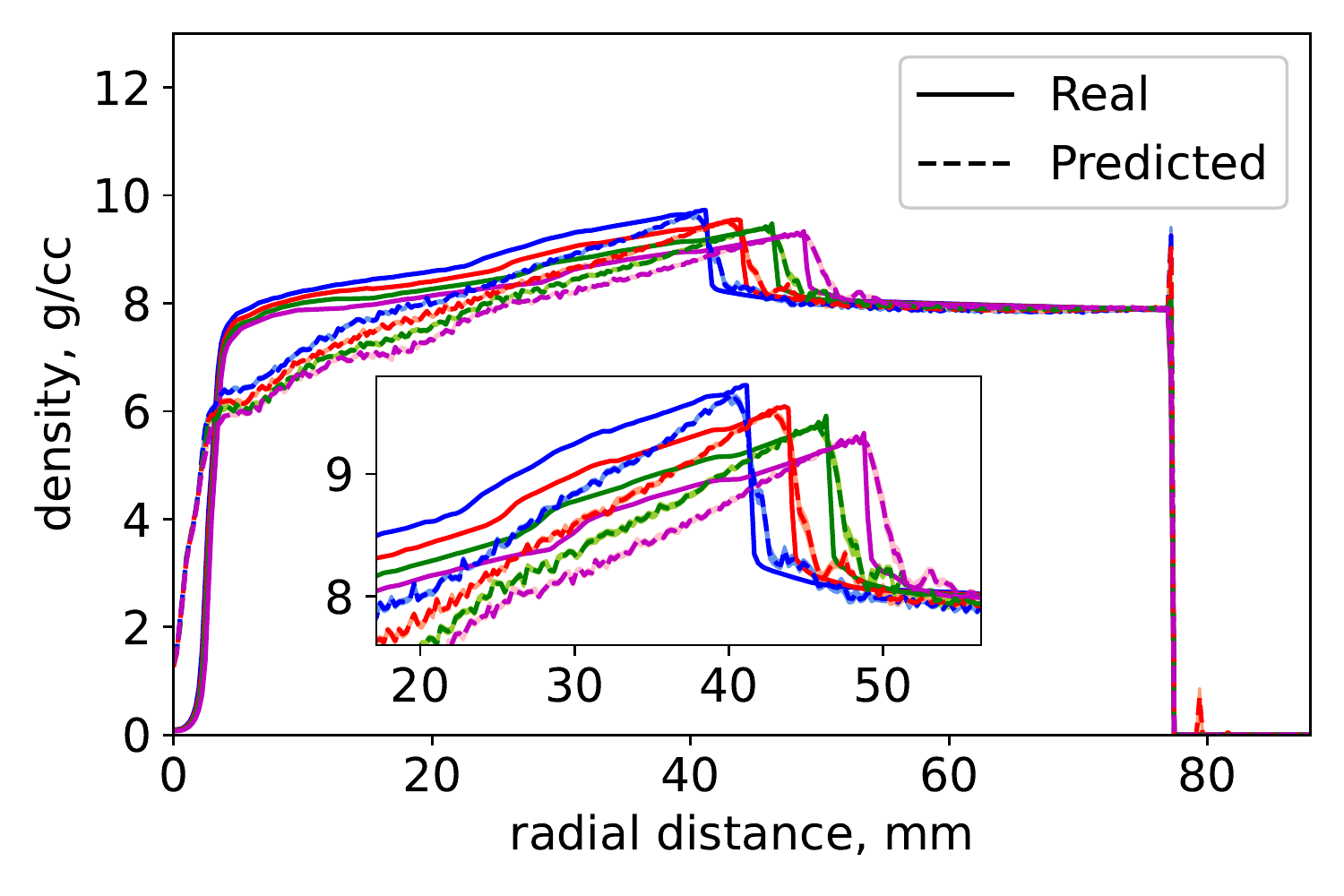} &
  \includegraphics[width=0.45\textwidth]{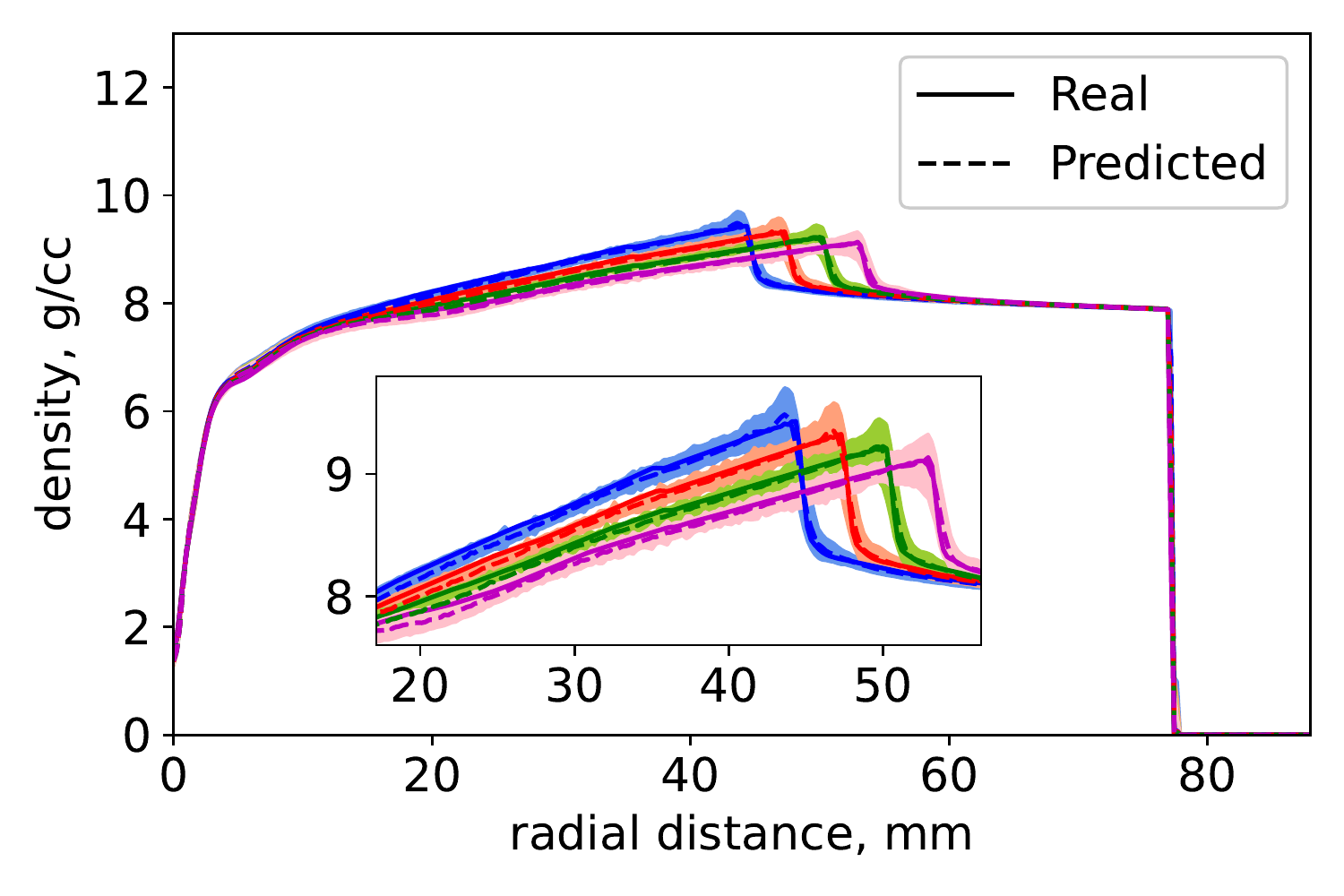}
  \\         
  (c) RMSE $=11.0\%$ &
  (d) RMSE $=8.98\%$
\end{tabular}
\caption{
    Inference on (a) J1,  
    (b) J8, 
    and (c) J11. 
    (d) Inference on J1 using network with $30\%$ dropout during testing.
    For all four cases, an ensemble of 1000 generated densities are plotted and are most visible in panels (a) and (d). The predicted result is an average of the 1000 generated samples.
}
\label{fig:mh_jen}
\end{figure}

Figure~\ref{fig:mh_jen} compares the generated density profiles 
with real snapshots from J1, J8, and J11. 
Simulation J1 is produced using the MG EOS and
EOS parameters within the 5D hypercube used to generate the training data. 
The corresponding density predicted by the cGAN-DF has 
low RMSE.
Here a predicted density is an average of a 1000 generated densities for given set of features.
Parameters associated with J8 lie outside the hypercube but 
the density is still captured accurately. 
The cGAN-DF is less robust to model mismatch in J11 as it fails to be 
as accurate as J1 or J8 when the MG EOS is replaced by a different model.

The introduction of the additional loss terms in eq.~\ref{eq:GANcomp_loss}
may steer the generator away from learning a true distribution match with $P$,
but as noted in \cite{isola2017image}, 
the L1 term in particular is necessary in order to produce convincing results.
During training, the cWGAN encounters multiple non unique feature locations 
as $F(\rho)$ are quantized into pixel width intervals.
In this context, an $L1$ loss would learn to fit a centroid of plausible $\rho$s
associated with the conditional input, 
resulting in blurry generated samples.
However, we observe that a larger $\lambda_{L1}$ weight results in sharper shocks in the samples generated by the cGAN-DF, 
in addition to improved texture matching.
This is a result of unique $F(\rho)$ values produced by the subpixel feature extraction method.
This difference in the nature of the training data and 
the additional constraints introduced by $L_M$ and $L_{\tilde F}$ in eq.~\ref{eq:GANcomp_loss}
limits the variation in the samples 
generated by the cGAN-DF for a given set of features. 
As in \cite{isola2017image}, we also observe that dropout during testing can increase the entropy in generated samples (see fig.~\ref{fig:mh_jen}(d)).

\subsection{Projection of Reconstructed Densities onto Hydrodynamic Manifold}
\label{sec:Projection of Reconstructed Densities onto Hydrodynamic Manifold}
We have utilized the robust hydrodynamic features, i.e., the shocks and edges in conjunction with  conditional GAN network architectures to learn the density sequences from a series of dynamic radiographs.  
While the density reconstructions appear to be satisfactory, there is no guarantee that these solutions will lie on a hydrodynamic manifold as given by the learned dynamic model.

Consequently, in this section we perform a projection of the density fields obtained from the learned model onto a hydrodynamic manifold by using parameter estimation. 
In this example we assume that we have a parametric form for the physics model, i.e., the MGR .  
Situations in which this is not the case may be addressed by learning a parametric model from the data or a neural network for example. \cite{kashiwa2010mggb,zhu2020generating,morawski2020neural}

Let $\Sigma$ denote the space of admissible parameters for our test problem.
Since we fix our initial conditions, each point $\sigma\in\Sigma$ is a vector of MG parameters $\sigma = (T_0, c_s, s_1, \Gamma_0, c_V)$. 
Given a numerical solution corresponding to the parameters $\sigma$ and an initial measurement time $t_0$ in some discrete set $T_{\text{data}}$, we can then extract a density sequence $\rho_n^{(\sigma)}(r)=\rho(r,t_0 + [n-1]\Delta t)$ from this solution that is contained in the hydrodynamic manifold $\mathcal{M}$. 
For an unknown density sequence $\rho_{\text{target}\,n}$ we can estimate its parameters $\sigma_{\text{target}}$ by solving the optimization problem
\begin{equation}
    \sigma_{\text{target}} = \argmin_{\sigma\in \Sigma_{\text{data}}}|| \rho_{\text{target}} - \rho_\sigma ||_2.
    \label{parameter_from_density_optimization}
\end{equation}
Here the combined $L_2$-norm over the $N_t=4$ time snapshots is defined as:
\begin{equation}
    ||\delta\rho||_2 := 
    \sqrt{\frac{1}{N_t V}
        \sum_{n=1}^{N_t}\sum_{i=1}^{N_x} (\delta \rho(r_i))^2 
        \Delta V_i
    },
    \label{eq:combined-L2norm}
\end{equation}
with $N_x$ being the total number of cells in radial direction, $\Delta V_i$ being the volume of an $i$-th cell, and $V$ the total volume.

This method of parameter estimation gives accurate results for an ideal case when the density profiles (a) lie on a restricted hydrodynamic manifold; (b) contain no noise; and (c) are known to be within the range of parameteric hypercube of the simulations database.
In such case, the accuracy of the parameter estimation is only limited by the error in interpolating between the neighboring entries in the database grid.
We may also use parameter estimation for the predictions of the density sequences obtained from our cWGAN and cGAN-DF networks.

The results of the manifold projection by parameter estimation are presented in Table~\ref{tab:params-estimation} for the test simulations J1-J10.
Recall that only the first four simulations J1-J4 are within the bounds of the database ($\pm10\%$ from the nominal values, see Table~\ref{tab:jennies}).

The three  columns in the table list the $L_2$ density mismatch between three density profiles: $\rho_{\rm GT}$ corresponds to the "ground truth", the true profile of the test; $\rho_{\rm GAN}$ is the GAN-reconstructed density profile, and $\rho_{\rm proj}$ represents the projection of the latter onto the dynamic manifold.
As can be seen from the table, the error of GAN reconstruction is comparable to the error of manifold reprojection.

\begin{figure}[htbp]
  \centering
  \begin{tabular}{cc}
  \includegraphics[width=0.60\textwidth]{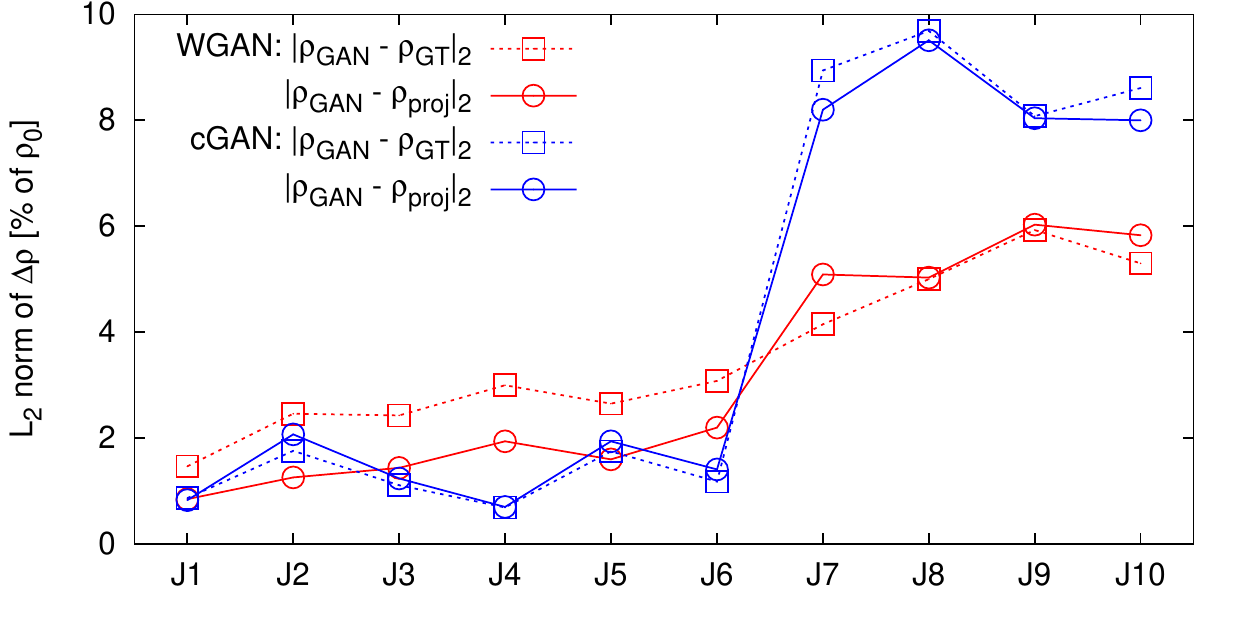} &
    \includegraphics[width=0.25\textwidth]{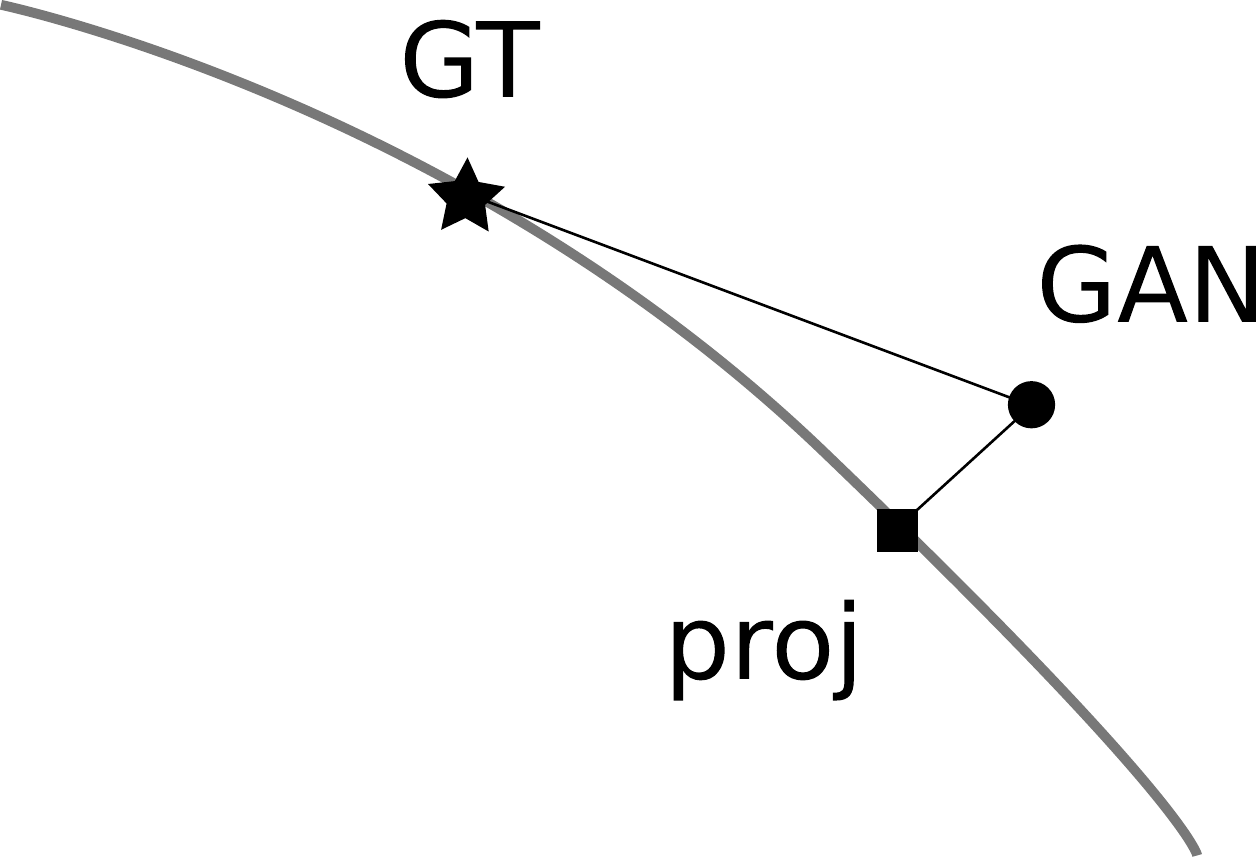}
  \end{tabular}
\caption{
Left: errors in density profiles for test simulations J1-J10.
Right: illustration of the $L_2$ distances between the ground 
truth (GT), GAN density reconstruction (GAN), and the projection 
of the GAN density profile (proj) onto the manifold (gray curve).
}
\label{fig:manifold-projection}
\end{figure}

\begin{table}[htp]
\begin{center}
\begin{adjustbox}{width=0.8\textwidth}
\begin{tabular}{l|ccc|ccc}
\hline
\hline
 & \multicolumn{3}{c}{cWGAN}
 & \multicolumn{3}{|c}{cGAN-DF}
\\
\hline
  & $\frac{||\rho_{\rm proj} - \rho_{\rm GT}||_2}{\rho_0}$
  & $\frac{||\rho_{\rm GT} - \rho_{\rm GAN}||_2}{\rho_0}$
  & $\frac{||\rho_{\rm GAN} - \rho_{\rm proj}||_2}{\rho_0}$
  & $\frac{||\rho_{\rm proj} - \rho_{\rm GT}||_2}{\rho_0}$
  & $\frac{||\rho_{\rm GT} - \rho_{\rm GAN}||_2}{\rho_0}$
  & $\frac{||\rho_{\rm GAN} - \rho_{\rm proj}||_2}{\rho_0}$
\\
\hline
\multicolumn{7}{l}{Minimization with L2 distance}
\\
\hline
J1 &   0.85 &  1.47 &  1.10  &   0.83 &  0.87 &  0.41\\
J2 &   1.26 &  2.46 &  1.19  &   2.07 &  1.76 &  0.54\\
J3 &   1.44 &  2.43 &  1.10  &   1.24 &  1.11 &  0.79\\
J4 &   1.94 &  3.00 &  1.22  &   0.70 &  0.69 &  0.65\\
J5 &   1.60 &  2.65 &  1.97  &   1.94 &  1.75 &  0.59\\
J6 &   2.20 &  3.08 &  1.65  &   1.41 &  1.18 &  0.60\\
J7 &   5.09 &  4.15 &  2.18  &   8.20 &  8.94 &  2.15\\
J8 &   5.03 &  5.00 &  1.08  &   9.51 &  9.69 &  1.99\\
J9 &   6.03 &  5.93 &  4.24  &   8.04 &  8.08 &  1.60\\
J10&   5.83 &  5.30 &  1.59  &   8.00 &  8.61 &  1.87\\
\hline
\multicolumn{7}{l}{Minimization with Wasserstein distance}
\\
\hline
J1  & 1.08 & 0.95 & 1.10 &  0.42 & 0.48 & 0.49 \\
J2  & 1.26 & 1.61 & 1.19 &  1.75 & 1.20 & 1.23 \\
J3  & 1.23 & 2.45 & 1.10 &  1.09 & 0.76 & 1.00 \\
J4  & 1.29 & 1.75 & 1.22 &  1.04 & 0.51 & 1.06 \\
J5  & 1.09 & 1.43 & 1.97 &  1.38 & 1.22 & 0.51 \\
J6  & 1.14 & 1.73 & 1.65 &  1.67 & 0.98 & 1.12 \\
J7  & 3.51 & 2.09 & 2.18 &  3.50 & 3.20 & 1.56 \\
J8  & 4.24 & 1.02 & 1.08 &  3.74 & 3.03 & 1.92 \\
J9  & 4.32 & 3.50 & 4.24 &  4.37 & 4.10 & 1.36 \\
J10 & 4.43 & 1.36 & 1.59 &  3.53 & 3.21 & 1.27 \\
\hline \hline
\end{tabular}
\end{adjustbox}
\end{center}
\caption{Errors in density profiles for projections 
onto hydrodynamic manifold by parameter estimation:
cWGAN (left) vs cGAN-DF (right).
All quantities are in terms of percentile deviations from the nominal values
(see Table~\ref{tab:nominal}), and $\rho_0 = 7.896\;\gcc$ is the nominal density.
} 
\label{tab:params-estimation}
\end{table}

\section{Discussion and Conclusion}
\label{sec:Discussion and Conclusion}

In this study, we consider a parametric study of a spherically symmetric shell implosion problem with subsequent expansion, focusing on the shock expansion phase, in an attempt to reconstruct complete density profiles using only the discontinuous hydrodynamic features: shock and edge in conjunction with learned dynamics so as to enable dynamic density reconstructions.
Our study is motivated by the fact that the features are normally much more easily identifiable in physical radiographs than removing scatter and other artifacts from the radiographic images, with the latter requiring solutions to the complex descattering,  beam physics, detector modeling, and inversion problems to recover the density fields.

We have demonstrated that a conditional generative adversarial network (cGAN) can be successfully applied to recover the density from a sequence of four positions of shock and edge, produced from corresponding time sequence of radiographs.
We used two different neural network architectures: conditional Wasserstein GAN (cWGAN, Section~\ref{sec:wgan}), and a cGAN with data fidelity (cGAN-DF, Section~\ref{sec:cgan}) and show that they produce results of comparable accuracy around 1\% in RMSE, for models which lie on the restricted hydrodynamic manifold.
For models which are beyond the boundaries of the hydrodynamic manifold, the networks do not perform as well (see Table~\ref{tab:params-estimation}), recovering density profiles with about 5\%-8\% accuracy.

Our study shows that the method of using dynamic behavior of radiographic features to reconstruct density is a viable approach, producing density profiles within the margin of error that is comparable or even better than obtained using the standard descattering + inversion algorithms.

\begin{figure}[htbp]
  \centering
  \includegraphics[width=0.60\textwidth]{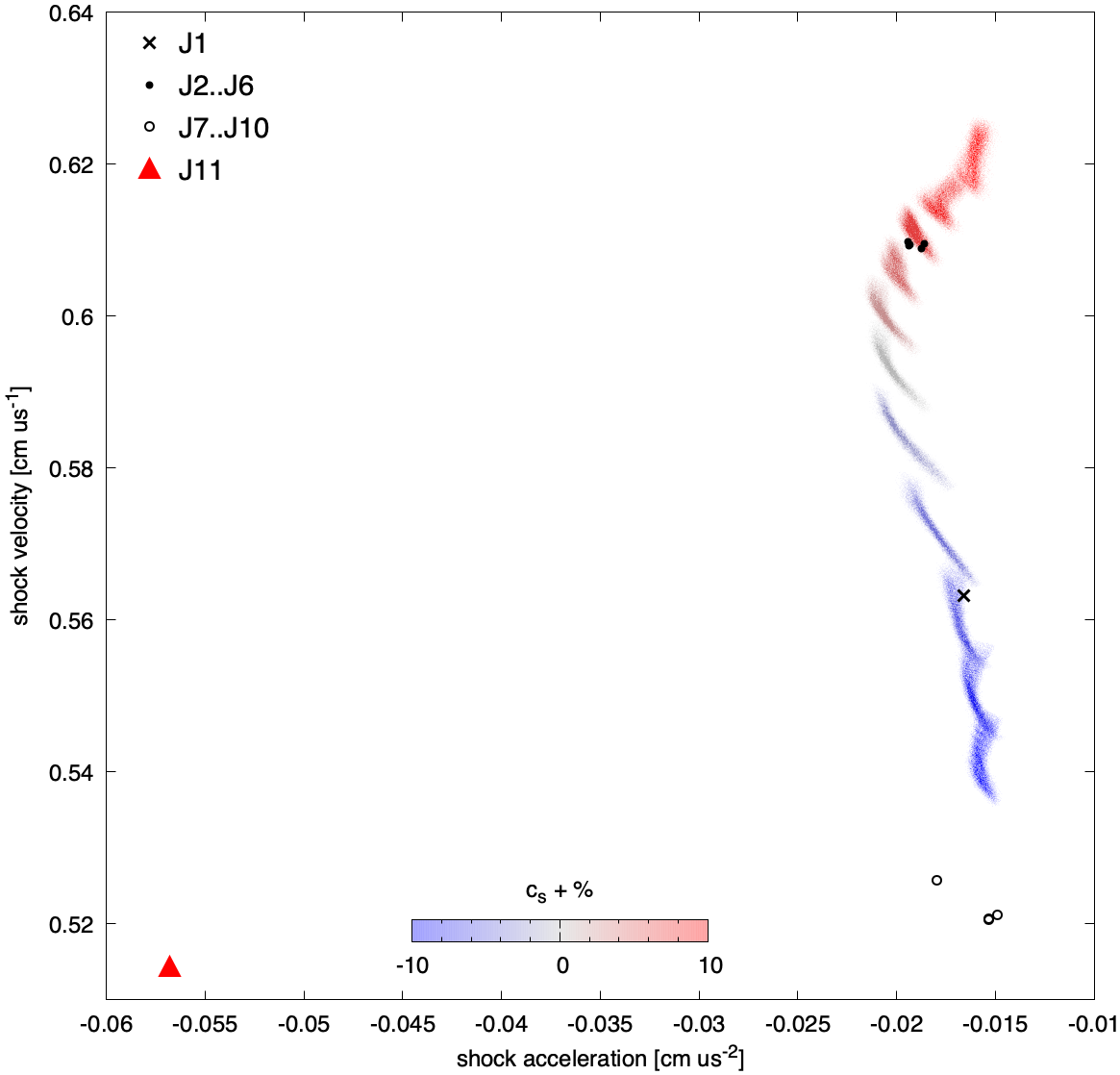}
  \caption{Map of the fitted velocity and acceleration of the shock for all 
  simulations in the Mie-Gr\"uneisen database, test simulations J1-J10 and a test 
  simulation J11 (red triangle) with a different equation of state 
  (using Sesame tables: see Table~\ref{tab:jennies}). 
  This plot illustrates why it may not be possible to meaningfully extrapolate 
  the hydrodynamic path from the simulation database.
  } 
  \label{fig:shock_vel_vs_acc}
\end{figure}
To better understand the properties of our underlying hydrodynamic simulations and also enable an understanding of the origin of the errors in the density reconstructions we examine the properties our hydrodynamic features, focusing on the velocity and acceleration of the shocks.
Figure~\ref{fig:shock_vel_vs_acc} depicts the point cloud of shock velocity and acceleration for all simulations in the database.
It also marks the location of the test simulations J1-J10. 
Because they were performed with the same MG equation of state, they should also lie on the hydrodynamic manifold.
One can clearly see from this visualization how J1-J6 are inside the point cloud, while J7-J10 (open circles) are not too far from it, in the sense that the data can be extrapolated to / resampled in that region to properly capture the shock behavior.
This figure also shows the location of J11 on the acceleration-velocity map (red triangle), which was performed with an entirely different equation of state.
It is clear that parameter estimation on this point can hardly produce meaningful results, because the point is too far away from the database.
This figure illustrates the difficulties encountered when extrapolation from the manifold is required.
Application of machine learning techniques cannot circumvent this problem, because  extrapolation will still be necessary.  However, to address this problem additional simulations using the same model that  better enable the training data to include the testing data may be performed.  Alternatively, alternative models may be utilized that are closer to the desired location in the shock velocity acceleration phase space.  This unique ability to examine the location of the observable features in the radiographs occurs to perform density reconstructions offers great potential in reducing density errors in reconstructions not available in the traditional dynamic radiographic reconstruction algorithms. 

Finally, we remark that our investigation of the nature of degeneracies of the density fields has revealed that while sub-pixel information appears to remove some degeneracies, given by the results indicated by the cGAN investigations and introduction of the L1 loss, uncertainties undoubtedly will be present due to the finite resolution of the imaging system.  
Consequently, the use of the GAN architecture is a necessary component of the learning of features to density fields. 


\begin{appendices}
\section{Subpixel Shock Extraction}
\label{sec:subpixel}

\begin{figure}[htbp]
  \centering
  \includegraphics[width=0.65\textwidth]{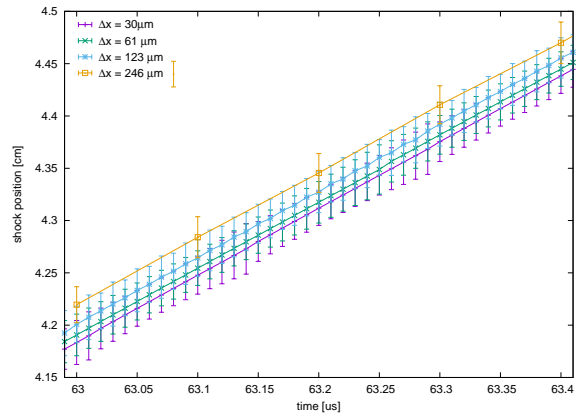}
  \caption{Extracted position of the shock for the test 
  problem with nominal parameter values, for different successively increasing 
  resolutions of the hydrodynamic numerical model. 
  The orange bar on the plot indicates the size of numerical cell for the coarsest
  resolution, demonstrating subpixel accuracy of shock and edge extraction.
  } 
  \label{fig:features-resolution}
\end{figure}

When recovering the shock position, it is convenient to use radial density $\mu(r) := 4\pi\rho r^2$ (with the units of $\gr\,\cm^{-1}$).
This quantity has an advantage of producing the mass directly when integrated over radius.
At a first step, we locate the shock and the grid cell with the maximum magnitude of the radial density gradient, $|\nabla_r\mu|$.
For the outgoing shock, this gradient is negative at the location of the shock and in a few neighboring cells.
In the remaining parts of the domain, the density does not drop very fast and because of the $O(r^2)$ term in the definition of $\mu$, the gradient of $\mu$ is positive almost everywhere.
A small connected segment $[r_{j-},r_{j+}]$ around the shock consisting of cells with the negative density gradient separates the domain into upstream and downstream parts.
We fit the radial density in the downstream and upstream $N$-cell neighborhood of the shock with linear regression, obtaining radial density approximations $\mu_-(r)$ and $\mu_+(r)$ to the left and to the right of the shock, respectively.
An accurate shock location can then be computed by assuming a discontinuous transition from $\mu_-(r)$ to $\mu_+(r)$ at some $r = r_s$ and solving the mass conservation equation for $r_s$:
\begin{equation}
    \int_{r_{j-}}^{r_s}\mu_-(r) dr + \int_{r_s}^{r_{j+}}\mu_+(r) dr = \sum_{j = j_-}^{j_+}\mu_j \Delta r.
\end{equation}
This again becomes a simple quadratic equation on $r_s$.
Clearly, the root must be picked such that ${r_s \in [r_{j-}, r_{j+}]}$.
Because of the imposed mass conservation, our technique is robust with respect to low-amplitude oscillating numerical artifacts near the shock.

Figure~\ref{fig:features-resolution} shows evolution of the extracted subpixel shock positions for different resolutions of hydrodynamic model.
This evolution demonstrates stable and convergent behavior of our subpixel feature extraction algorithm.

\end{appendices}


\begin{backmatter}

\bmsection{Acknowledgments}
The authors thank Luke Pfister and Jonah~M. Miller for their helpful feedback.
This work used resources provided by the LANL Institutional Computing Program. 
LANL is operated by Triad National Security, LLC, for the National Nuclear Security Administration of the U.S. DOE (Contract No. 89233218CNA000001).
This work was also partially supported by NSF grant number CCF-1763896.
This work is authorized for unlimited release under LA-UR-21-31230.

\bmsection{Disclosures}
The authors declare no conflicts of interest.

\bmsection{Data availability} 
Data underlying the results presented in this paper are not publicly available at this time but may be obtained from the authors upon reasonable request.

\end{backmatter}


\end{document}